\documentclass{article}
\usepackage[utf8]{inputenc}

\usepackage{lipsum}
\usepackage{amsfonts}
\usepackage{graphicx}
\usepackage{epstopdf}
\usepackage{algorithmic}
\ifpdf
  \DeclareGraphicsExtensions{.eps,.pdf,.png,.jpg}
\else
  \DeclareGraphicsExtensions{.eps}
\fi
\usepackage{amsmath}
\usepackage{amsthm}
\usepackage{float}
\usepackage{color}
\usepackage{mathtools}
\usepackage{caption}
\usepackage{subcaption}
\usepackage[margin=0.75in]{geometry}

\makeatletter
\newcommand{\sublabel}[1]{\protected@edef\@currentlabel{\thefigure(\thesubfigure)}\label{#1}}
\makeatother

\title{Comparing Finite-Time Lyapunov Exponents and Lagrangian Descriptors for identifying phase space structures in a simple two-dimensional, time-periodic double-gyre model}

\author{By: Timothy Getscher\thanks{Research conducted at: United States Naval Academy, Annapolis, MD. Current address: MIT-WHOI Joint Program in Oceanography, Woods Hole, MA (tgetscher@gmail.com)}\\ Sponsor: Kevin McIlhany, Ph.D. \thanks{United States Naval Academy, Annapolis, MD (mcilhany@usna.edu)}}

\date{March 12, 2021}

\begin{document}

\maketitle

\begin{abstract}
This paper compares the advantages, limitations, and computational considerations of using Finite-Time Lyapunov Exponents (FTLEs) and Lagrangian Descriptors (LDs) as tools for identifying barriers and mechanisms of fluid transport in two-dimensional time-periodic flows. These barriers and mechanisms of transport are often referred to as ``Lagrangian Coherent Structures,'' though this term often changes meaning depending on the author or context. This paper will specifically focus on using FTLEs and LDs to identify stable and unstable manifolds of hyperbolic stagnation points, and the Kolmogorov-Arnold-Moser (KAM) tori associated with elliptic stagnation points. The background and theory behind both methods and their associated phase space structures will be presented, and then examples of FTLEs and LDs will be shown based on a simple, periodic, time-dependent double-gyre toy model with varying parameters.
\end{abstract}

\section{Introduction}

\hspace{\parindent} In the past several decades, the dynamical systems approach to Lagrangian transport has been applied to a variety of problems in computational fluid dynamics (CFD), such as transport and mixing in oceanic flows \cite{coulliette2001} \cite{mancho2006} \cite{mendoza2014} \cite{rypina2009}, and microfluidic mixers \cite{mcilhany2011} \cite{ottino2004}.  More recently, techniques from the field have also been applied to chemical reaction dynamics \cite{katsanikas2020} \cite{naik2019}. In this paper, we motivate the dynamical systems approach to Lagrangian transport by comparing two numerical diagnostics which reveal phase space structures in two-dimensional, time-periodic incompressible flows.  These geometric structures will be shown to constrain or facilitate transport in flow fields. Specifically, we seek to identify stable and unstable manifolds of stagnation points with hyperbolic stability, and the Kolmogorov-Arnold-Moser (KAM) tori of stagnation points with elliptic stability. \\

In literature, the term ``Lagrangian Coherent Structure'' (LCS) often has different meanings depending on the author or paper. It is most often referred to broadly and qualitatively, as a structure which organizes the rest of the flow into ordered patterns \cite{haller2015} \cite{haller2000} . Haller and Yuan define LCS in \cite{haller2000} as material lines for which \textit{hyperbolicity times} (which they have defined) attain local extrema. Haller also provides an alternate definition in \cite{haller2001}, where LCS are defined as the local maxima of the Finite-Time Lyapunov Exponent (FTLE) field. A similar definition of LCS was given in \cite{shadden2005}, which defines Lagrangian Coherent Structures as the ridges of FTLE fields. In this paper, we use the term broadly, perhaps synonymous with ``phase space structure,'' as a way to categorize flow fields into regions of distinct behavior. More often, however, we will simply refer to particular structures, such as stable and unstable manifolds or KAM invariant tori.  \\

There are many numerical diagnostics that have been used to qualitatively explore transport and mixing in flow fields, such as FTLEs \cite{shadden2005}, Lagrangian Descriptors (LDs) \cite{mancho2013b} \cite{mendoza2010b}, Poincare maps \cite{dumas2014}, and Encounter Volume \cite{rypina2017}, to name a few. This paper will specifically compare FTLEs and LDs. These two methods reveal similar patterns in flow fields, and they are both used to identify phase space structures such as stable and unstable manifolds or KAM tori. Despite the similarity of the two diagnostics, a comparison of the methods is useful since they have different limitations, advantages, and computational considerations. \\

In Section \ref{sec:Gyre}, a simple model of a two-dimensional time-dependent double-gyre will be described, which will be used as our example to compute FTLE and LD fields. A double-gyre is a large-scale feature of ocean circulation, typically occurring in the mid-latitudes. Specifically, the term double-gyre refers to two ``gyres,'' or circulations, which occur adjacent to each other and flow in opposite directions, such as a system containing both a subpolar and subtropical gyre, for example. Inter-gyre transport remains an active and important area of research within the study of general ocean circulation \cite{burkholder2011} \cite{coulliette2001} \cite{foukal2016} \cite{levang2020} \cite{liu1994}. In this paper, we use a simple two-dimensional toy representation of this phenomenon to illustrate concepts. \\

In Section \ref{sec:FTLE}, the definition and computation of FTLEs will be described, and their relationship to the stable and unstable manifolds of stagnation points with hyperbolic stability will be discussed. Similarly, in Section \ref{sec:LD}, the definition and computation of LDs and their relationship to stable and unstable manifolds will be shown, along with a discussion of their relationship to KAM tori. Lastly, in Section \ref{sec:results}, a number of examples of LDs, FTLEs, and Poincare Maps will be shown using the time-dependent double-gyre subject to varying parameters.

\section{The Double-Gyre} \label{sec:Gyre}

\hspace{\parindent}  We will now introduce a simple example of double-gyre flow, which will be used to compute the numerical diagnostics that will be shown in Sections \ref{sec:FTLE} - \ref{sec:results}. The system we describe is prevalent in literature and is commonly used for various topics in fluid dynamics \cite{shadden2005}. Suppose that a fluid is two-dimensional, incompressible, and inviscid. Then the velocity field can be obtained from the derivatives of a scalar function, $\psi$, called the streamfunction: 

\begin{align}
    u &= -\frac{\partial \psi}{\partial x} \label{eqn:psi}\text{,} \\
    v &= \frac{\partial \psi}{\partial y} \text{.} \nonumber
\end{align}

In the context of dynamical systems, Equation \eqref{eqn:psi} describes a time-dependent Hamiltonian vector field, for which $\psi$ represents the Hamiltonian function. In this paper we will sometimes use the vector $\vec{u} = (u,v)$ to denote velocity and $\vec{x} = (x,y)$ to denote spatial coordinates.  \\

Consider the following ``general'' equation for a time-dependent double-gyre system: 

\begin{eqnarray}
\label{eqn:UPsi}
\psi &=& A \sin(\pi f(x,t,\epsilon,\omega))\sin(\pi y) \text{,}
\end{eqnarray}

\noindent which we will define on the domain $(x,y) \in [0~ 2] \times [0~ 1]$. In this system $f$ should be chosen such that $f(x = 0,t,\epsilon,\omega) = 0$ and  $f(x = 2,t,\epsilon,\omega) = 2$. This function $f$ determines the time-dependent behavior of the system, where $\epsilon$ and $\omega$ will be used to represent the ``amplitude'' and ``frequency'' of a periodic perturbation, respectively. The parameter $A$ is a constant which scales the Hamiltonian (and therefore the magnitude of the velocity field). The first thing we will note is that when $f(x,t,\epsilon,\omega) = x$, the streamfunction describes a system which is identical for all time, and so it is called ``steady'' (i.e. time-\textit{independent}). In this case, the streamfunction and its corresponding velocity field reduce to: 

\begin{align}
\psi &= A\sin(\pi x) \sin(\pi y) \label{eq:psisteady}\text{,} \\
u &= -\frac{\partial \psi}{\partial x} = -A\pi\sin(\pi x)\cos(\pi y) \text{,} \nonumber \\
v &=  \frac{\partial \psi}{\partial y} = A\pi\cos(\pi x)\sin(\pi y) \text{.} \nonumber
\end{align} \\

\begin{figure}[H]
\centering
\includegraphics[width=.6\textwidth]{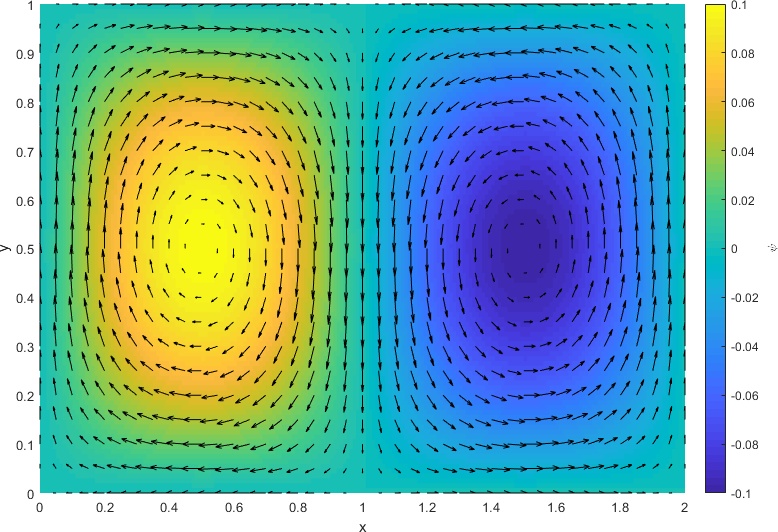}
\caption{Velocity field $\vec{u}(x,y)$ overlaid on top of the contour plot of $\psi$ for the steady-state double-gyre system described by Equation \eqref{eq:psisteady}}.
\label{fig:vel_field}
\end{figure}

The streamfunction and corresponding velocity field for this system have been shown in Figure \ref{fig:vel_field}. This system has several notable features. First, there is no flux through the boundaries or between gyres, which is to say that $u = 0$ along the lines $x = 0$, $x = 1$, and $x = 2$, and that $v = 0$ along the lines $y = 0$ and $y = 1$. The line $x = 1$ is often called a ``separatrix'' since it has the effect of separating the two gyres. There are also several stagnation points in the flow field. \textit{Stagnation points} are any point $(x,y)$ for which $\vec{u}(x,y) = 0$. For this flow system, stagnation points exist at all corners, between the two gyres along the vertical boundaries  (i.e. $(x,y) = (1,0), (1,1)$), and at the center of either gyre (i.e. $(x,y) = (0.5, 0.5), (1.5,0.5)$ ). \\

Stagnation points in Hamiltonian systems are characterized by linearized stability, the nature of which is determined by finding the eigenvalues of the Jacobian of the linearized velocity field when evaluated at the stagnation point. The linearized behavior of these stagnation points can have either Hyperbolic or Elliptic stability (note that this reduction only applies to Hamiltonian systems, such as the one we use). \textit{Hyperbolic} stability occurs when the eigenvalues of the Jacobian evaluated at a stagnation point are both pure real and have opposite signs. \textit{Elliptic} stability occurs when these eigenvalues are pure imaginary. The behavior of hyperbolic and elliptic trajectories near these stagnation points will be discussed more in Sections \ref{sec:FTLE} and \ref{sec:LD}.

Next, we will consider a case where the double-gyre is unsteady (i.e. time-\textit{dependent}). A sinusoidal time-dependence is often used, described by: 

\begin{eqnarray}
f(x,t,\epsilon,\omega) &=& a(t)x^2+b(t)x\text{,} \label{eqn:timedep} \\
a(t,\epsilon,\omega) &=& \epsilon \sin(\omega t)\text{,} \nonumber \\
b(t,\epsilon,\omega) &=& 1 - 2 \epsilon \sin(\omega t)\text{.} \nonumber
\end{eqnarray}

Note that $f(x,t,\epsilon,\omega)$ could be replaced with some other function, but the sinusoidal time-dependence is useful since it is both simple and periodic (with period $T = 2\pi/\omega$). \\

\begin{figure}[H]
\centering
\includegraphics[width=0.5\linewidth]{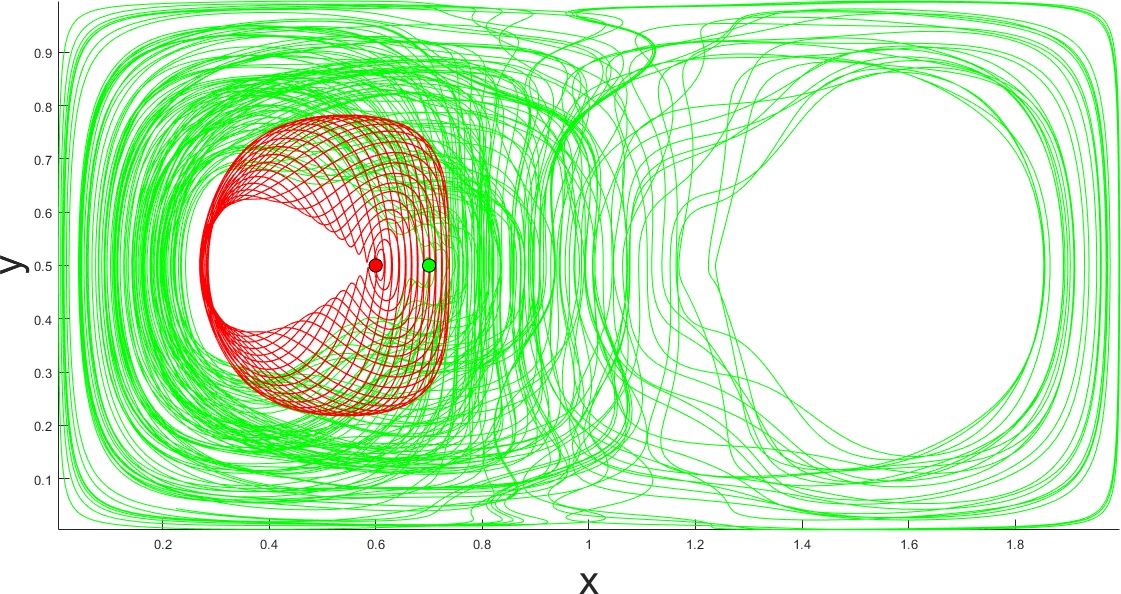}
\caption{Two trajectories in the time-dependent double-gyre such as in Equations \eqref{eqn:UPsi} and \eqref{eqn:timedep} initialized at $t_0 = 0$ with parameters $A = .1$, $\epsilon = .25$ and $\omega = 2$. The trajectories are calculated until $t = 1500$. The red trajectory remains inside of left gyre for all time, while the green trajectory moves between the two gyres several times. The dots at $(x,y) = (.6,~.5)$ (trajectory shown in red) and $(x,y) = (.7,~.5)$ (trajectory shown in green) represent the initial conditions. }
\label{fig:twotrajs}
\end{figure}

With this unsteady version of the double-gyre, one can verify that, much like the steady double-gyre, there is no flux through the boundaries (i.e., $u = 0$ along the lines $x = 0$ and $x = 2$, and $v = 0$ for $y = 0$ and $y = 1$). However, what we have previously called the separatrix appears to oscillate left and right, to a minimum $x = 1-\epsilon$ and a maximum of $x = 1+\epsilon$. At this point, this line is no longer a separatrix, since it is no longer a barrier to transport (a \textit{separatrix} is most often defined as a Lagrangian feature which acts as a barrier to transport between two flow regimes). In Figure \ref{fig:twotrajs}, two trajectories initialized at $t = 0$ in the double-gyre with parameters $A = .1$, $\epsilon = 0.25$ and $\omega = 2$ are shown, up to time $t = 1500$. Note that one trajectory has stayed ``inside'' of its gyre, and the other has moved between the left and right-hand sides of the domain several times. By ``moving between gyres'' we mean that the parcel crosses the center line for which $u = 0$ (which was previously at $x = 1$ but is now oscillating). The mechanisms which act as barriers to transport for such fluid parcels (and similarly, those which facilitate transport) will be discussed further in Sections \ref{sec:FTLE} and \ref{sec:LD}. 

As the line for which $u = 0$ moves, so do the stagnation points on its upper and lower boundaries. Similarly, the stagnation points which begin at the center of both gyres begin oscillating back and forth. These stagnation points can now be called ``instantaneous stagnation points'' (ISPs). An \textit{instantaneous stagnation point} is a point in space occurring at any instant in time for which $\vec{u}(x,y,t) = 0$. Now that the basic behavior and features of the time-dependent double-gyre have been described, we will use it to perform the numerical diagnostics that are the focus of this paper.  

\section{Finite-Time Lyapunov Exponents} \label{sec:4}
 \label{sec:FTLE}
 
\hspace{\parindent} The first diagnostic that we will consider is the Finite-Time Lyapunov Exponent (FTLE). We have mentioned that we will be using diagnostics to look for ``phase space structures.'' The term \textit{phase space} may have slightly different meanings depending on context. In ordinary differential equations, and in our case, we use phase space to refer to the space of dependent variables. For the two-dimensional flow that we consider, phase space is synonymous with the physical space in which the fluid flows, i.e. the space with coordinates $(x,y)$.  \\

\subsection{Definition and Computation}

\hspace{\parindent} The concept of \textit{Lyapunov Exponents} began in the late 1800s, when Aleksandr Lyapunov addressed the problem of stability in dynamical systems. Lyapunov Exponents are the exponential separation rate of infinitesimally close trajectories \cite{katok1995}. Since Lyapunov Exponents are defined only in infinite time, the Finite-Time Lyapunov Exponent (FTLE) is often used in computation. Conceptually, the FTLE is a time average of the maximum separation rate for a pair of particles advected in a flow, after some time interval. To constructs FTLEs, the right Cauchy-Green deformation tensor is computed for an initial condition and a grid of its neighboring points. The Cauchy-Green deformation tensor, in this case, is a $2x2$ matrix which measures local stretching between points, and is constructed as in Equation \eqref{eqn:CG}:

\begin{equation}
    \Delta = \frac{\text{d} \phi_t^{t+\tau}(\vec{x})}{\text{d} \vec{x}}^* \frac{\text{d} \phi_t^{t+\tau}(\vec{x})}{\text{d} \vec{x}}\text{,} \label{eqn:CG}
\end{equation}

\noindent where $\phi_t^{t+\tau}$ is the flow map of the dynamical system, meaning that it indicates the final location of a trajectory beginning at $\vec{x} = (x,y)$ and time $t$ after it is integrated to $t+\tau$. Numerically, we compute $\frac{\text{d} \phi_t^{t+T}(\vec{x})}{\text{d} \vec{x}}$ in finite-difference form such as in Equation \eqref{eqn:differencing}:

\begin{equation}
\label{eqn:differencing}
    \left.\frac{\text{d} \phi_t^{t+\tau}(\vec{x})}{\text{d} \vec{x}}\right\vert_{x_{i,j}} = 
    \begin{pmatrix}
        \frac{x_{i+1,j}(t+\tau) - x_{i-1,j}(t+\tau)}{x_{i+1,j}(t) - x_{i-1,j}(t)} & \frac{x_{i,j+1}(t+\tau) - x_{i,j-1}(t+\tau)}{y_{i,j+1}(t) - y_{i,j-1}(t)} \\
        \\
        \frac{y_{i+1,j}(t+\tau) - y_{i-1,j}(t+\tau)}{x_{i+1,j}(t) - x_{i-1,j}(t)} & \frac{y_{i,j+1}(t+\tau) - y_{i,j-1}(t+\tau)}{y_{i,j+1}(t) - y_{i,j-1}(t)}
    \end{pmatrix}\text{,}
\end{equation}

\noindent where $x_{i,j}(t)$ indicates the x-location of the particle in grid position $i,j$ at time $t$, etc. The FTLE is considered a ``forward'' FTLE when $\tau$ is positive and a ``backward'' FTLE when $\tau$ is negative.\\

We can then compute the FTLE of a trajectory using: 

\begin{equation}
    \sigma_{t}^\tau(\vec{x}) = \frac{1}{|\tau|}\ln{\sqrt{\lambda_\text{max}(\Delta)}}\text{,}
\end{equation}

\noindent where $\lambda_\text{max}(\Delta)$ is the maximum eigenvalue of the deformation tensor, $\Delta$. This process is repeated for each point on a grid of initial conditions to construct an ``FTLE field,'' i.e. a scalar field in $\mathbb{R}^2$ for which the value of the field at $(x,y)$ refers to the FTLE value of a trajectory beginning at $(x,y)$ after some finite integration time, $\tau$.\\

We note that FTLEs are, by definition, a differentiated quantity. Any student who has been asked to compute the derivative of a signal has likely found that differentiation has the tendency to add noise to computations. For FTLEs, one consequence of this is that computed FTLE values are highly sensitive to the grid spacing of available initial conditions. A simple experiment can illustrate this dilemma. Consider the coarse grid of initial conditions (and their corresponding FTLE field) shown in Figure \ref{fig:FTLEcoarse}, for which trajectories are known on the interval $[t_0 ~t_f] = [0 ~12]$. For this grid, FTLE values can be computed for all interior grid points. In Figure \ref{fig:FTLEfine}, a slightly denser grid is shown.  As expected, FTLE values from the original grid have changed, since  trajectories of new ``closer'' initial conditions were used to compute FTLE values. As the grid spacing continues to decrease, the FTLE value will continue to change (and should converge to the analytic FTLE value for the initial condition and specified integration time). Also note that even if trajectories of grid points along the boundaries are known, FTLE values cannot be computed on the edges of the domain, since to do so would require trajectories which lie outside of the domain. 

\begin{figure}[H]
\begin{subfigure}[t]{0.5\textwidth}
\includegraphics[width=\textwidth]{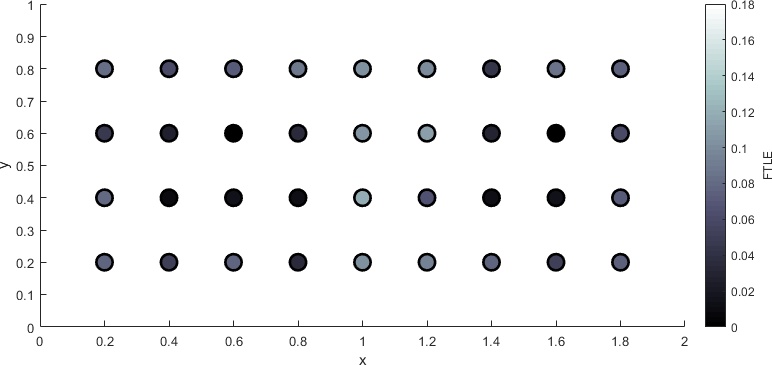}
\caption{}
\sublabel{fig:FTLEcoarse}
\end{subfigure}
\begin{subfigure}[t]{0.5\textwidth}
\includegraphics[width=\textwidth]{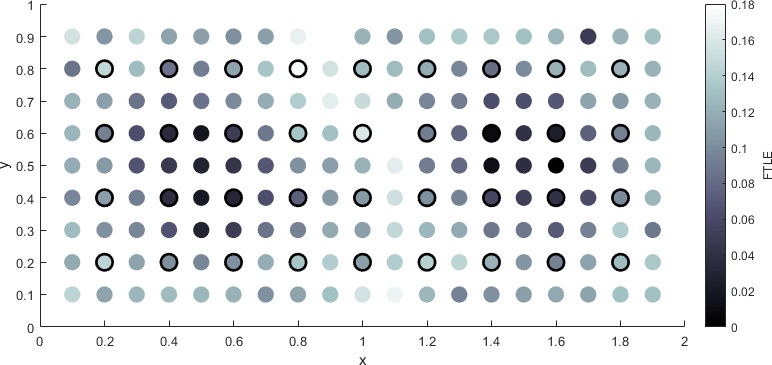}\caption{}
\sublabel{fig:FTLEfine}
\end{subfigure}
\caption{Two FTLE Fields for the double-gyre such as in Equations \eqref{eqn:UPsi} and \eqref{eqn:timedep} with parameters $A = .1$, $\epsilon = .1$, $\omega = 2$ and $\tau = 12$. Panel (a) uses a coarse grid of initial conditions that begins as 11x6 (note that FTLE values can only be computed for the interior 9x4 grid, since FTLEs cannot be computed along the boundaries).  Panel (b) uses a slightly finer 21x11 grid of initial conditions. The grid points which remain the same between the two cases have been outlined by a black edge in panel (b). Note that since the ``neighboring'' initial conditions and their corresponding trajectories have changed in panel (b), that the FTLE values at the ``old'' locations have changed in response. }
\end{figure}

\begin{figure}[H]
\begin{subfigure}[t]{0.5\textwidth}
\includegraphics[width=\textwidth]{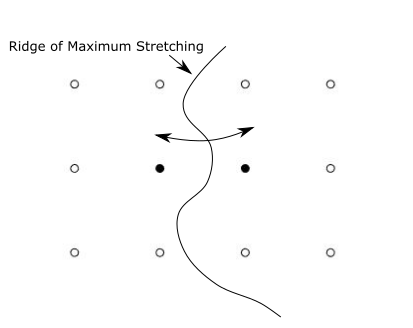}
\caption{}
\sublabel{fig:FTLEGridGood}
\end{subfigure}
\begin{subfigure}[t]{0.5\textwidth}
\includegraphics[width=\textwidth]{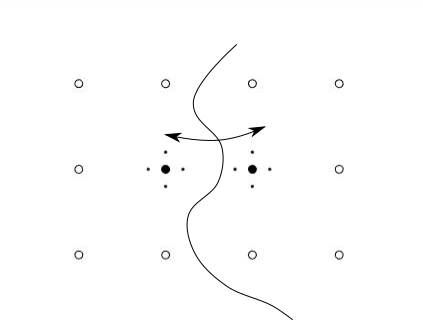}
\caption{}
\sublabel{fig:FTLEGridBad}
\end{subfigure}
\caption{An example grid of initial conditions is shown. We wish to calculate FTLEs for the initial conditions that are filled in black. In panel (a), the shown grid alone is used for the FTLE computation. In panel (b), a new grid of initial conditions, beginning very close to the center grid point, is generated for the computation. The case in panel (b) is intuitive for many people, but should be avoided, since the ridge of maximal stretching might not be captured by any FTLE values.}
\label{fig:FTLEGrid}
\end{figure}

After reading the previous discussion, a student who is new to computing FTLEs might do the following: after creating the grid for which one wants to compute FTLEs, compute four new trajectories with initial conditions that are very, very close to the original grid point. This would, in fact, more closely approximate the analytic FTLE for that initial condition. However, doing so for a grid of initial conditions could be extremely misleading. This is because when computing FTLEs, we are usually less interested in the exact FLTE value at a specific location, but rather we are interested in roughly identifying ridges of maximal stretching (these ridges are associated with structures in phase space and will be discussed in the next subsection). If one wishes to compute FTLE values at two locations that are on either side of a ridge of maximal stretching, then using new initial conditions that are too close to the original grid points can  ``hide'' this ridge. To visualize this, consider the grid and ridge of maximal stretching in Figure \ref{fig:FTLEGrid}. In panel \ref{fig:FTLEGridGood}, the two points on either side of the ridge are used for the computation. In this situation, the stretching caused by the ridge will be reflected in the computed FTLE values, since at least one pair of points will experience stretching associated with the ridge. In panel \ref{fig:FTLEGridBad}, four initial conditions surrounding each grid point have been generated in order to compute FTLEs. Notice that both sets of points are contained on either side of the ridge. It can be seen that two points which both begin on one side of a ridge will not experience the same extent of ``stretching'' from each other that would be experienced by two points on either side of the same ridge. In the case for which virtual trajectories were computed, the ridge of stretching will likely not be reflected in any FTLE values. Therefore, it is only in the first situation that the desired phase space structure has been identified.  

\begin{figure}[H]
\begin{subfigure}[t]{0.48\textwidth}
\includegraphics[width=\textwidth]{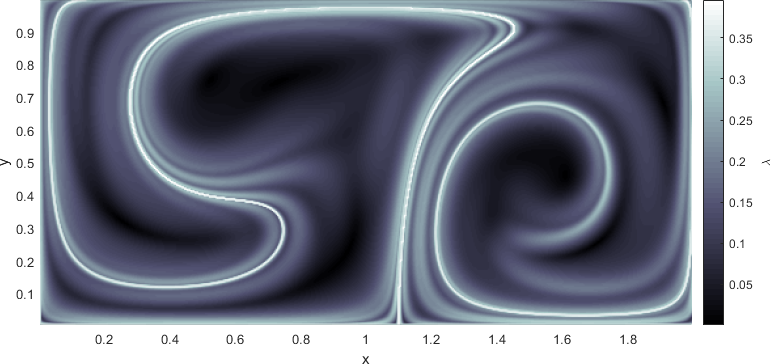}
\end{subfigure}\hfill
\begin{subfigure}[t]{0.48\textwidth}
\includegraphics[width=\textwidth]{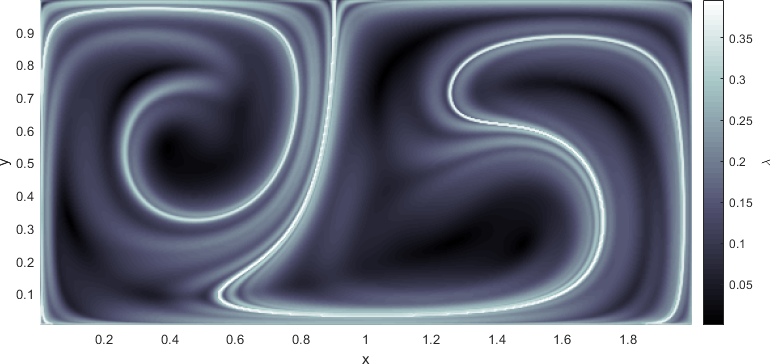}
\sublabel{fig:backward}
\end{subfigure}\hfill
\caption{FTLE fields for the double-gyre such as in Equations \eqref{eqn:UPsi} and \eqref{eqn:timedep} with parameters $A = .1$, $\epsilon = .25$, $\omega = .5$, and $\tau = 15$. Forward FTLEs are shown in panel (a) and backward FTLEs are shown in panel (b). Note the distinct ridges of maximal stretching that have formed.}
\label{fig:FTLEfwdbwd}
\end{figure}

An example of two FTLE fields (one forward and one backward) are shown in Figure \ref{fig:FTLEfwdbwd}. One sees that the ridges of maximal stretching develop in the FTLE fields. These ridges approximate the stable and unstable manifolds of hyperbolic stagnation points in the flow field. 

\subsection{FTLEs for observing Stable and Unstable Manifolds}

\begin{figure}[H]
    \centering
    \includegraphics[width = .4\textwidth]{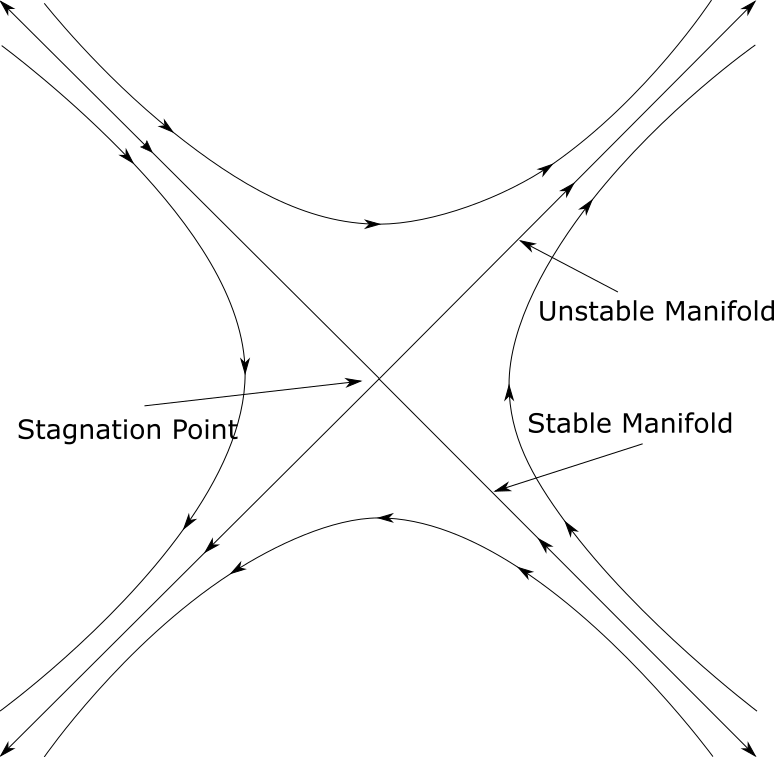}
    \caption{A hyperbolic stagnation point, with its associated stable and unstable manifolds also labeled.}
    \label{fig:HyperbolicFP}
\end{figure}

\hspace{\parindent} We previously defined stagnation points and instantaneous stagnation points in the context of dynamical systems. These stagnation points can exhibit several types of stability which characterize nearby trajectories. We are currently interested in stagnation points with hyperbolic stability, which means that the eigenvalues of the Jacobian evaluated at the stagnation point have real values with opposite sign. Stagnation points with hyperbolic stability possess stable and unstable manifolds. The \textit{stable manifold} of a hyperbolic stagnation point is defined as the curve for which trajectories lying on the curve will exponentially decay towards a hyperbolic point as time increases, and similarly, an \textit{unstable manifold} is a curve whose trajectories will exponentially grow away from a hyperbolic stagnation point as time increases. These manifolds and their hyperbolic stagnation point may move or oscillate in time. In Figure \ref{fig:HyperbolicFP}, a simple schematic of a hyperbolic stagnation point is shown at the intersection of a stable and unstable manifold. 

One can see in Figure \ref{fig:HyperbolicFP} that all initial conditions on either side of the stable manifold, for instance, will be advected away from the stable manifold in forward time, so that the stable manifold acts as a boundary which separates parcels that move exponentially away from each other. This means that the maximal stretching of a flow field in forward time often occurs near a stable manifold, and so the ridges of forward FTLE fields correspond to stable manifolds (and similarly, the ridges of backward FTLEs will correspond to unstable manifolds). In literature, FTLE fields are widely used to approximate stable and unstable manifolds for autonomous as well as both periodically and aperiodically forced systems, and many references can be found which conduct such analyses \cite{branicki2010} \cite{haller2001} \cite{rypina2009} \cite{shadden2005}. \\

\begin{figure}[H]
\begin{subfigure}[t]{0.46\textwidth}
\includegraphics[width=\textwidth]{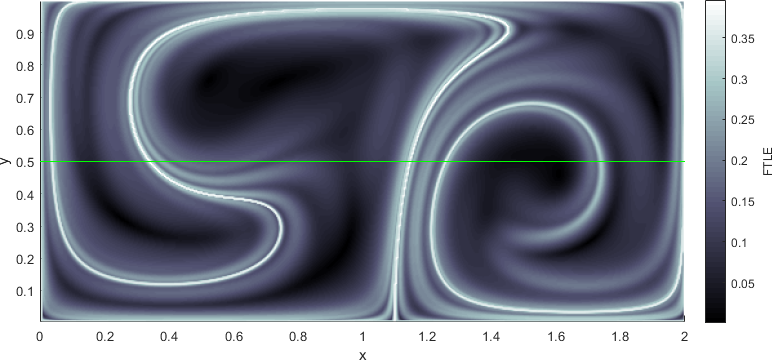}
\caption{}
\sublabel{fig:FTLEtau15}
\end{subfigure}\hfill
\begin{subfigure}[t]{0.46\textwidth}
\includegraphics[width=\textwidth]{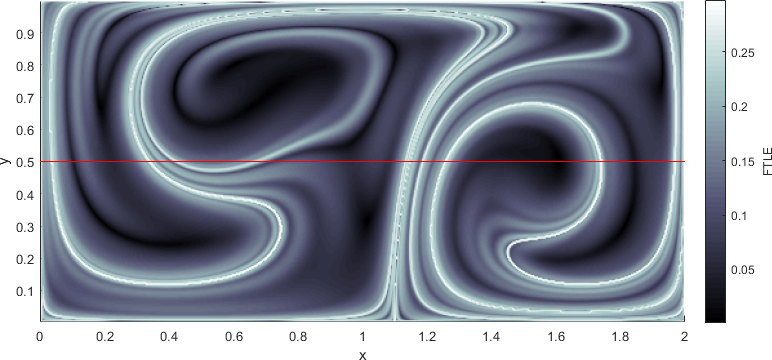}
\caption{}
\sublabel{fig:FTLEtau20}
\end{subfigure}
\centering
\begin{subfigure}[t]{0.48\textwidth}
\includegraphics[width=\textwidth]{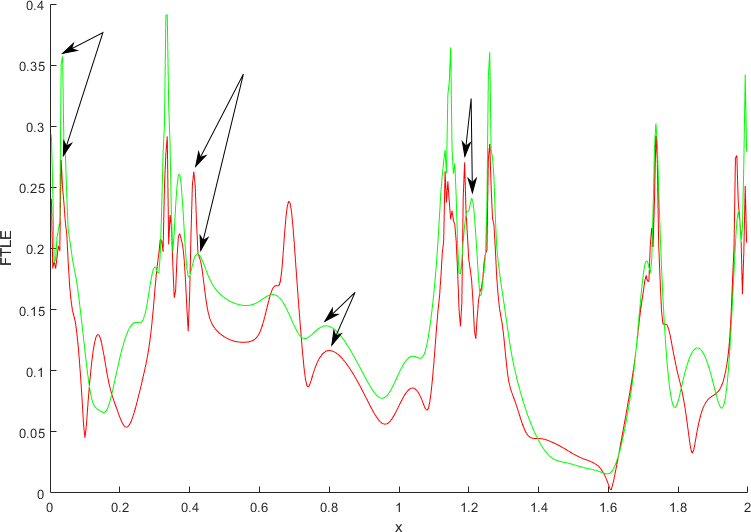}
\caption{}
\sublabel{fig:FTLEsingular}
\end{subfigure}
\caption{Forward FTLE fields for the double-gyre such as in Equations \eqref{eqn:UPsi} and \eqref{eqn:timedep} with parameters $A = .1$, $\epsilon = .25$, and $\omega = .5$ are shown. The case of $\tau = 15$ is shown in panel (a) and the case of $\tau = 20$ is shown in panel (b). In panel (c), the FTLE value for cross-sections along the line $y = 0.5$ are shown (the green curve corresponds to $\tau = 15$ and red curve corresponds to $\tau = 20$.). Note that some ridge locations have changed slightly between the two cases.}
\label{fig:FTLEchangetau}
\end{figure}

We would like to know how varying integration time affects the structure of FTLE fields. Figure \ref{fig:FTLEchangetau} shows FTLE fields for the same set of initial conditions, where the only parameter varied is integration time. In panel \ref{fig:FTLEtau15} the integration time is $\tau = 15$ and in panel \ref{fig:FTLEtau20} the integration time is $\tau = 20$. In panel \ref{fig:FTLEsingular}, cross-sections of the FTLE field have been taken at $y = 0.5$ for both the $\tau = 15$ and $\tau = 20$ cases from the previous panels. As integration time increases, longer manifold segments (and therefore more complexity) can be seen in the field. More peaks can then be seen in panel (c) for the case of longer $\tau$. However, we notice that some peaks have changed $x$-locations between the two cases. \\

The Forward and Backward FTLE fields of a system can be summed together, in order to visualize the approximations of stable and unstable manifolds simultaneously. These curves intersect to form ``lobes'' which are bounded on one side by a stable manifold and another side by an unstable manifold. Since stable and unstable manifolds are material curves, all trajectories which begin inside of a lobe will remain bounded by the stable and unstable manifolds for all time, and so they are ``stuck'' in the lobe \cite{coulliette2001}. A \textit{material curve} is a surface of one less dimension than the flow domain that is made up entirely of particle trajectories. In systems such as ours, the lobes will often ``stretch'' and ``fold'' so that they become incorporated into both gyres. In this way, lobes made by the intersection of stable and unstable manifolds act as the mechanism for transport in the time-dependent double-gyre. This process is often referred to as ``lobe dynamics'' and has been described in many papers such as \cite{coulliette2001} \cite{raynal2006}.

\section{Lagrangian Descriptors} \label{sec:LD}
 
\hspace{\parindent} The next diagnostic we will use to identify phase space structures in the double-gyre system are Lagrangian Descriptors (LDs). LDs are another trajectory-based scalar diagnostic that were first described by Mendoza and Mancho as a tool for finding hyperbolic trajectories in dynamical systems \cite{mendoza2010b}. They were originally defined as the Euclidean arc-length of a trajectory, but have since been extended to include various $p$-norm quantities, and other quantities such as the integrals of acceleration or curvature \cite{mancho2013b}. In this paper we will consider only trajectory arc-length, which is the most common definition of LDs. 

\subsection{Definition and Computation}

\hspace{\parindent} Lagrangian Descriptors, according to the definition we will use, are a measure of the Euclidean arc-length of trajectories. We will use the function $M$ to denote the LD value for an initial condition with location $\vec{x}$ beginning at time $t_0$ and with integration window $\tau$.  In integral form (where the integrand is the magnitude of the velocity field), $M$ is written:

\begin{equation}
    M(\vec{x},t_0)  = \int_{t_0 - \tau}^{t_0 + \tau}{||\vec{u}}(\vec{x}(t),t)||~dt \label{eqn:Mfunction} \text{,}
\end{equation}

\noindent or in the discrete case, we will use the function $MD$ to denote discrete Lagrangian Descriptor values on a finite grid: 

\begin{align}
    MD_{i} = \sum_{n = -N}^{N} \sqrt{(x_{i}^{n+1}-x_{i}^n)^2 + (y_{i}^{n+1}-y_{i}^n)^2}\text{,} \label{eqn:MD}
\end{align}

\noindent where $n$ indexes the current time-step, $i$ indexes a particular initial condition, $N$ indicates the number of time steps in the integration window, and $x_i^{n+1}$ and $y_i^{n+1}$ depend on the initial conditions and flow map of the system. Note that $t = n\Delta t$, $\tau = N \Delta t$. \\

One might also wish to separate Lagrangian Descriptors into two parts, one that is integrated forward in time and one that is integrated backward in time. This will help us later to separately identify stable and unstable manifolds, similar to the use of forward and backward FTLEs. This is done with the separated equations

\begin{align}
    M(\vec{x},t_0) &= M^f(\vec{x},t_0) + M^b(\vec{x},t_0)\text{,} \label{eqn:Mfunction}\\
    M^f(\vec{x},t_0) &= \int_{t_0}^{t_0 + \tau}{||\vec{\mathbf{u}}}(\vec{x}(t),t)||~dt \label{eqn:MfunctionFWD} \text{,}\\
    M^b(\vec{x},t_0)  &= \int_{t_0-\tau}^{t_0}{||\vec{\mathbf{u}}}(\vec{x}(t),t)||~dt\text{.} \label{eqn:MfunctionBWD} 
\end{align}

We note that Lagrangian Descriptors are, by definition, an integrated quantity. Any student who has been asked to integrate a signal numerically has likely found that integration has a smoothing effect. \\

One useful feature of $M$ is that its value is not dependent on neighboring trajectories. To visualize this, consider the coarse grid of initial conditions and their corresponding LD fields in Figure \ref{fig:LDcoarse}, and the slightly finer grid in Figure \ref{fig:LDfine}. Note that LD values from initial conditions that exist in both cases are identical. No change is observed because the computation of LDs does not require information from neighboring points. 

\begin{figure}[H]
\begin{subfigure}[t]{0.5\textwidth}
\includegraphics[width=\textwidth]{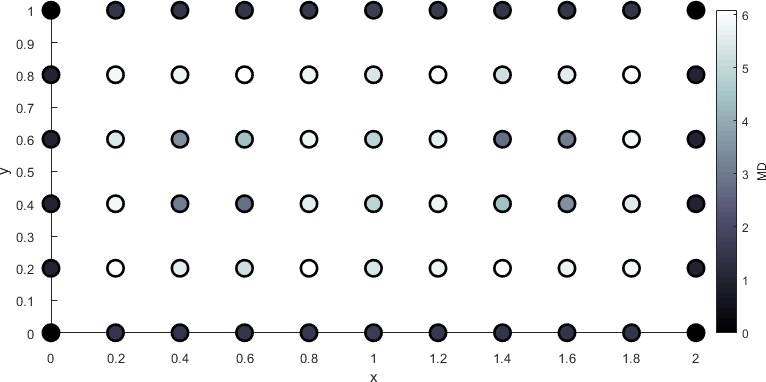}
\caption{}
\sublabel{fig:LDcoarse}
\end{subfigure}
\begin{subfigure}[t]{0.5\textwidth}
\includegraphics[width=\textwidth]{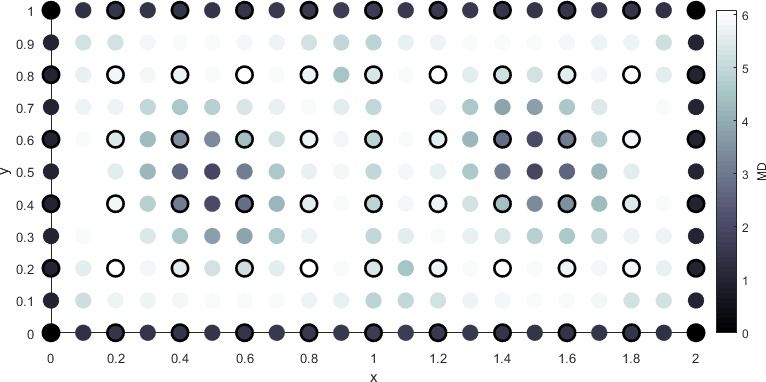}
\caption{}
\sublabel{fig:LDfine}
\end{subfigure}
\caption{LD values for the double-gyre such as in Equations \eqref{eqn:UPsi} and \eqref{eqn:timedep} with parameters $A = .1$, $\omega = .5$, and $\tau = 12$. In panel (a) a coarse grid of 11x6 points is used, and in panel (b) a slightly finer grid of 21x11 points is used. Initial conditions which remain the same in both cases have been shown outlined in black in panel (b). Note that the value of $M$ at these locations remains identical in both cases. Also note that LD values can be computed along the boundaries, so long as the trajectory is known.}
\end{figure}

\subsection{Lagrangian Descriptors for observing Stable and Unstable Manifolds} 

\hspace{\parindent} When computing LDs for a grid of initial positions, one begins to identify features which are very similar to the FTLE ridges shown in the previous section. Figures \ref{fig:forwardLD} and \ref{fig:backwardLD} contain forward and backward LD fields for the double-gyre with parameters $\epsilon = 0.25$ and $\omega = .5$, with integration time $\tau = 15$ (note that these are the same parameters used in the FTLE fields of Figure \ref{fig:FTLEfwdbwd}). These again represent the stable and unstable manifolds of stagnation points.\\ 

\begin{figure}[H]
\begin{subfigure}[b]{0.48\textwidth}
\includegraphics[width=\textwidth]{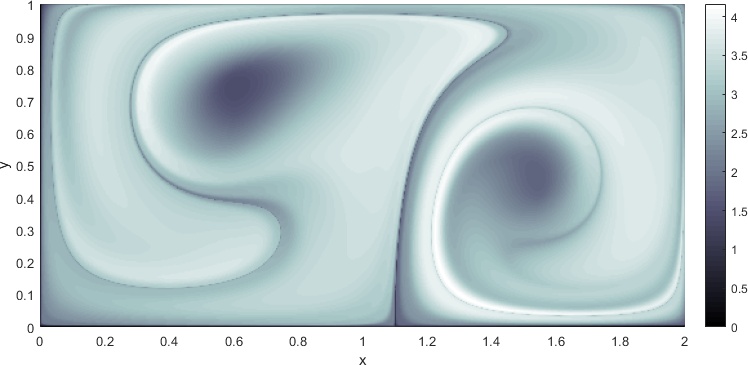}
\caption{}
\sublabel{fig:forwardLD}
\end{subfigure}\hfill
\begin{subfigure}[b]{0.48\textwidth}
\includegraphics[width=\textwidth]{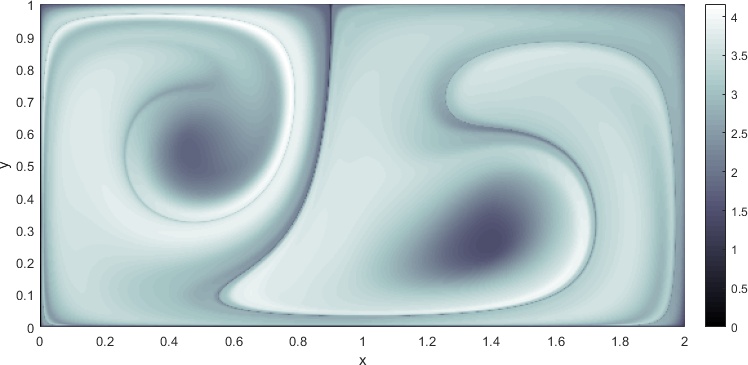}
\caption{}
\sublabel{fig:backwardLD}
\end{subfigure}\hfill
\caption{Forward (a) and Backward (b) LD fields for the double-gyre such as in Equations \eqref{eqn:UPsi} and \eqref{eqn:timedep} with parameters $A = .1$, $\epsilon = 0.25$, $\omega = .5$, and $\tau = 15$, computed using Equations \eqref{eqn:MfunctionFWD} and \eqref{eqn:MfunctionBWD}.}
\end{figure}

There exists a rigorous mathematical connection between the stable and unstable manifolds displayed in LD fields and the exact stable and unstable manifolds of hyperbolic stagnation points. This result was first proved by Lopesino et al. for two-dimensional autonomous and nonautonomous flows \cite{lopesino2017}. The proof was next extended to 3D dynamical systems in Garc\'ia-Garrido et al. \cite{garcia-garrido2018}, and more recently it has been shown for the stable and unstable manifolds of normally hyperbolic invariant manifolds in Hamiltonian systems with two or more degrees of freedom in Naik et al. \cite{naik2019}. These proofs will not be discussed here, but we will instead mention the heuristic argument presented, for instance, by Mancho et al. \cite{mancho2013b}.  \\

\begin{figure}[H]
\begin{subfigure}[t]{0.48\textwidth}
\includegraphics[width=\textwidth]{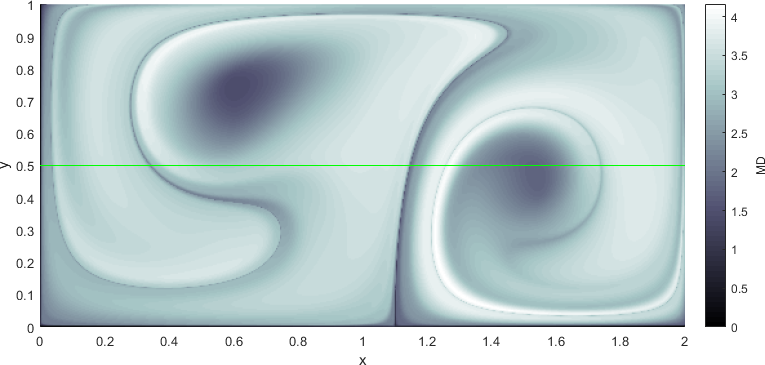}
\caption{}
\sublabel{fig:LDtau15}
\end{subfigure}\hfill
\begin{subfigure}[t]{0.48\textwidth}
\includegraphics[width=\textwidth]{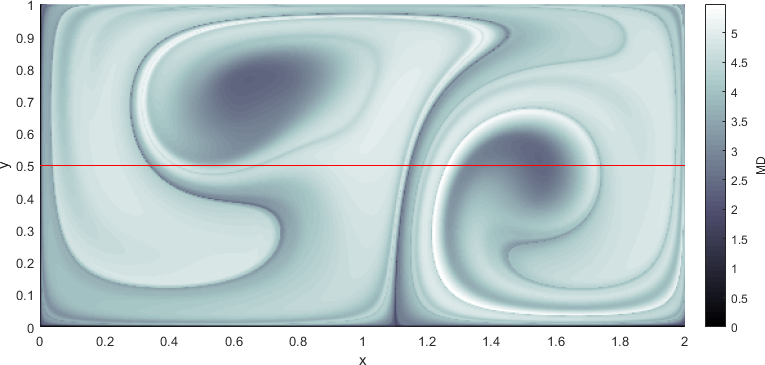}
\caption{}
\sublabel{fig:LDtau20}
\end{subfigure}
\hfill
\centering
\begin{subfigure}[t]{0.48\textwidth}
\includegraphics[width=\textwidth]{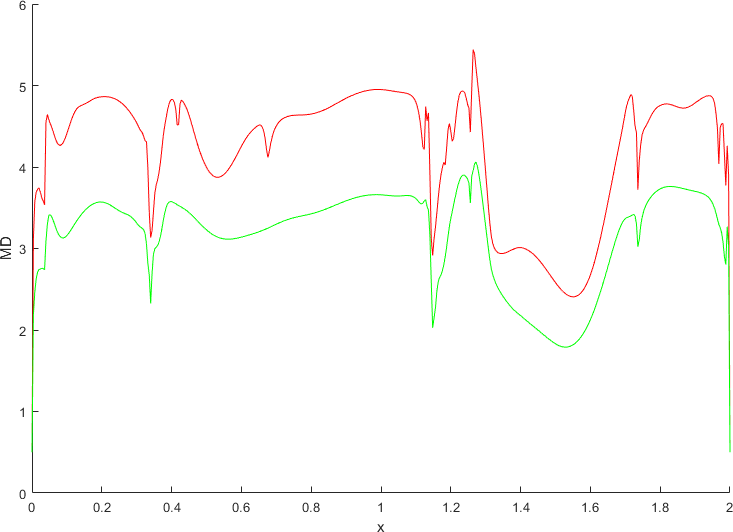}
\caption{}
\sublabel{fig:LDRidge}
\end{subfigure}
\caption{Forward LD fields for the double-gyre such as in Equations \eqref{eqn:UPsi} and \eqref{eqn:timedep} with parameters $A = .1$, $\epsilon = 0.25$, and $\omega = .5$. The case of $\tau = 15$ is shown in panel (a) and the case of $\tau = 20$ is shown in panel (b). In panel (c), the value of $MD$ from a cross section at $y = 0.5$ in each case is shown as a function of $x$ (green curve corresponds to $\tau = 15$ and red curve corresponds to $\tau = 20$). In panel (c), discontinuities in LD values from the shorter (green) simulation remain in the longer (red) simulation, illustrating that the location of stable or unstable manifolds do not change as $\tau$ increases. However, more discontinuities begin to appear in the long (red) curve, showing that increasing $\tau$ for LDs increases the length of the resolved manifolds.} 
\label{fig:SingularFeature}
\end{figure}

The heuristic argument is as follows: trajectories which begin ``near'' each other and remain so throughout their time evolution are expected to have arc-lengths which are ``close.'' However, at the boundaries between regions consisting of trajectories with qualitatively different behavior, we expect that the arc-lengths of trajectories on either side of this boundary will not be ``close.'' Instead, these boundaries are marked by an ``abrupt change'' in $M$, meaning that the derivative of $M$ transverse to these boundaries is discontinuous. This abrupt change occurs near regions separated by the stable and unstable manifolds of hyperbolic trajectories. Figure \ref{fig:SingularFeature} demonstrates this numerically with $MD$ for a case of the double-gyre with $\epsilon = .25$ and $\omega = 0.5$. The values of $MD$ along the line $y = 0.5$ are plotted, and one can see that a discontinuity in the derivatives of $MD$ occur at the locations of stable and unstable manifolds. Note that these ``discontinuities'' are not precisely the singular features described in \cite{mancho2013}, since the shown cross-section is not necessarily transverse to the stable and unstable manifolds. The method, however, is useful for visualizing abrupt changes in $MD$. \\

Similar to FTLE fields, we would like to know how varying integration time affects the structure of LD fields. LD fields exhibit a similar behavior to FTLEs in that as $\tau$ is increased, longer manifold segments and increased complexity can be seen in the domain. However, the locations of the perceived stable and unstable manifolds do not change when $\tau$ is increased. This can be understood first through the more rigorous connection between LDs and stable/unstable manifolds in \cite{lopesino2017}, \cite{garcia-garrido2018}, and \cite{naik2019}, and is also an observed fact for those who compute LDs. Consider the following two fields of LDs, one with an integration time of $\tau = 15$ in Figure \ref{fig:LDtau15}, and one with an integration time of $\tau = 20$ in Figure \ref{fig:LDtau20}. It can be seen in the $\tau = 20$ case that ridge locations have not changed compared to $\tau = 15$, but have only gained complexity. This can also be seen in panel \ref{fig:LDRidge}, since the discontinuities in $MD$ occur at the same location in both cross-sections. \\

\subsection{Lagrangian Descriptors and KAM Tori}

\hspace{\parindent} Now that some examples of Lagrangian Descriptors have been shown (with stable and unstable manifolds clearly visible), one can see that there are ``other'' features in the LD fields that do not correspond with manifolds. Specifically large dips tend to appear near the center of gyres. These dips correspond to stable KAM tori. In this section, we will describe the relationship between Lagrangian Descriptors and KAM tori. First, Kolmogorov-Arnold-Moser (KAM) theory will be briefly introduced. The concept of Poincare Maps will be introduced, and some examples of Poincare Maps for our double-gyre model will be shown. Finally, the relationship between Lagrangian Descriptors and KAM invariant tori will be described. 

\subsubsection{KAM Theory}

\hspace{\parindent} KAM theory arises from elliptic trajectories that are affected by a small, time-periodic perturbation. By elliptic trajectories, we mean trajectories which begin near a stagnation point with elliptic stability. A stagnation point has elliptic stability when the Jacobian of the velocity field evaluated at the stagnation point are pure imaginary. \\

For a conservative flow in an unperturbed system, the streamfunction of the flow, $\psi(x,y)$, can be transformed in terms of the action-angle variables, $(I,\theta)$ by replacing $\psi(x,y)$ with the new Hamiltonian, $H(I)$. The details of this transformation will not be described here but are common in classical mechanics. The equations of motion in the transformed system become:

\begin{eqnarray}
\frac{\text{d} I}{\text{d} t} = -\frac{\partial H}{\partial \theta} &=& 0\text{,} \\
\frac{\text{d} \theta}{\text{d} t} = ~~\frac{\partial H}{\partial I} &=& \omega(I)\text{.}
\end{eqnarray}

After integrating these equations, one sees that $I$ simply indexes particle trajectories, and that the motion of trajectories is periodic with angular frequency $\omega(I)$. These trajectories can be viewed as lying on curves, or ``tori'' within phase space. When the system is perturbed, almost all tori are preserved while others are destroyed. This is the main result of the Kolmogorov-Arnold-Moser (KAM) theorem. More specifically, the KAM theorem maintains that after a sufficiently small perturbation, the tori of most trajectories are preserved, except for those which have an angular frequency that is rationally related to the frequency of perturbation \cite{dumas2014}. Which is to say: 

\begin{equation}
    \alpha = \frac{\omega(I)}{\omega_0} = \frac{p}{q}\text{,}
\end{equation}

\noindent where $(p,q)$ are integers, and $\omega_0$ is the frequency of perturbation.\\

The trajectories for which the ratio $\alpha$ is sufficiently irrational (meaning their angular frequencies are not rationally related to the frequency of perturbation and that they do not lie very close to one of these trajectories) are associated with structures called KAM invariant tori. These tori will remain invariant under perturbation and are barriers to transport --- meaning any orbits beginning ``inside'' of the tori will remain inside throughout the system's dynamic evolution \cite{dumas2014}. Meanwhile, tori with rational $\alpha$ and their close neighbors will be ``destroyed,'' or in other words, they will not densely fill out a curve. One way to easily view these tori is through a tool known as a Poincare Map.

\subsubsection{Poincare Maps} \label{subsec:Poincare}
\begin{figure}[H]
    \centering
    \begin{subfigure}[t]{0.48\textwidth}
    \includegraphics[width=\textwidth]{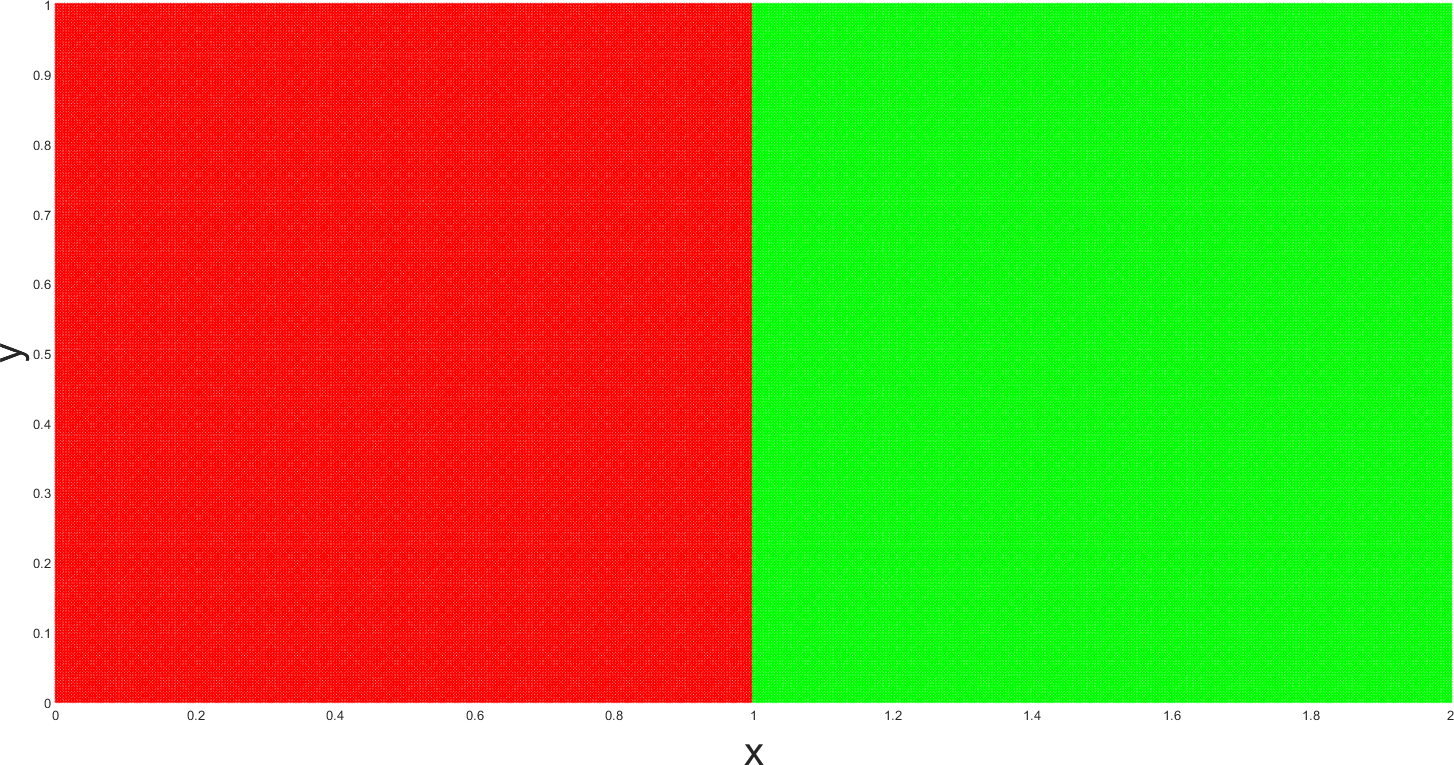}
    \caption{}
    \sublabel{fig:SimulationIC}
    \end{subfigure}
    \begin{subfigure}[t]{0.48\textwidth}
    \includegraphics[width=\textwidth]{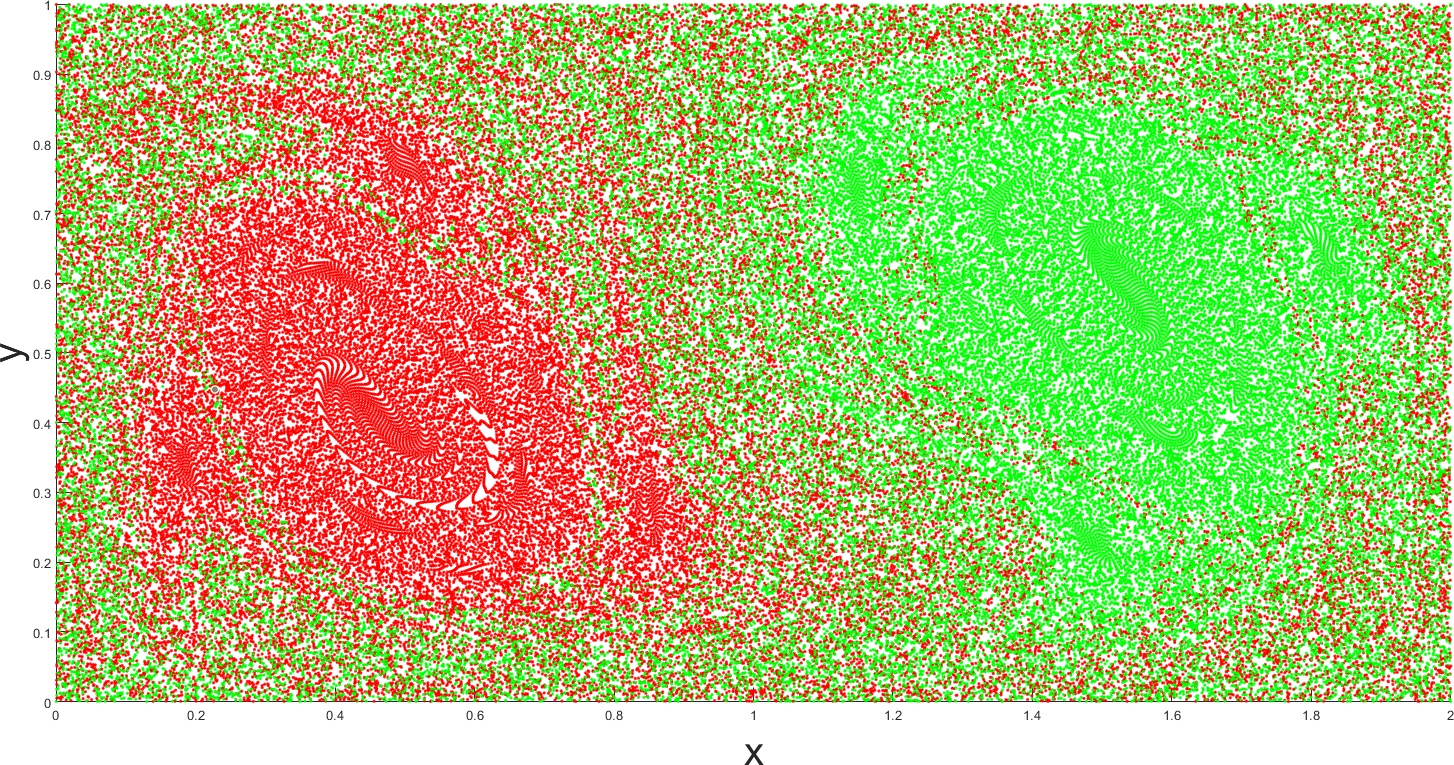}
    \caption{}
    \sublabel{fig:SimulationF}
    \end{subfigure}
    \caption{Initial conditions (a) and final state (b) of particles advected in double-gyre flow such as in Equations \eqref{eqn:UPsi} and \eqref{eqn:timedep} with parameters $\epsilon = 0.25$ and $\omega = 2$, after $\tau = 100.53$ (32 iterations of the period $T = 2\pi/\omega$). The lack of mixing around the center of the gyres is related to the invariant tori from the Poincare Map in Figure \ref{fig:Poincare}}
    \label{fig:Simulation}
\end{figure}

\hspace{\parindent} Recall Figure \ref{fig:twotrajs} for which two trajectories were released. One moved chaotically around the domain, and the other remained on one side of the domain for the entirety of the simulation. This calls for a tool to determine whether certain trajectories are ``trapped'' permanently in regions of the domain (particularly near elliptic centers). Consider Figure \ref{fig:Simulation}, which contains the initial and final positions of fluid parcels after being advected for 32 iterations of the period of perturbation of the system.  From this simulation alone, however, one cannot determine whether a parcel is actually stuck in one of these islands (perhaps some left and returned, or will escape at a later time if the simulation is continued). Drawing inspiration from KAM theory allows us to develop Poincare Maps that helps us determine whether some trajectories will be ``caught'' in regions of elliptic stability.  

A \textit{Poincare Map} is a type of recurrence map which is constructed by plotting the positions of trajectories at integer multiples of the period of perturbation, $T= 2 \pi / \omega$ \cite{dumas2014}. Figure \ref{fig:PoincareF} contains a Poincare Map of the double-gyre with parameters $\epsilon = 0.25$, $\omega = 2$, and integrated for 4000 multiples of the period of perturbation. The initial conditions used to create this map are shown in Figure \ref{fig:PoincareIC}. Note that when constructing these maps, a much less dense grid of initial conditions is used compared to a grid one might prefer when running a simple simulation of advected parcels. This is because too many initial conditions makes it more difficult to observe distinct features in Poincare Maps. Typically, one wishes to see solid or dashed curves which one can interpret as stable or unstable tori. In Figure \ref{fig:PoincareF}, evidence of both resonant and non-resonant tori can be seen when beginning with the coarse grid of parcels in Figure \ref{fig:PoincareIC}.  \\

\begin{figure}[H]
\centering
    \begin{subfigure}[t]{0.48\textwidth}
    \includegraphics[width=\textwidth]{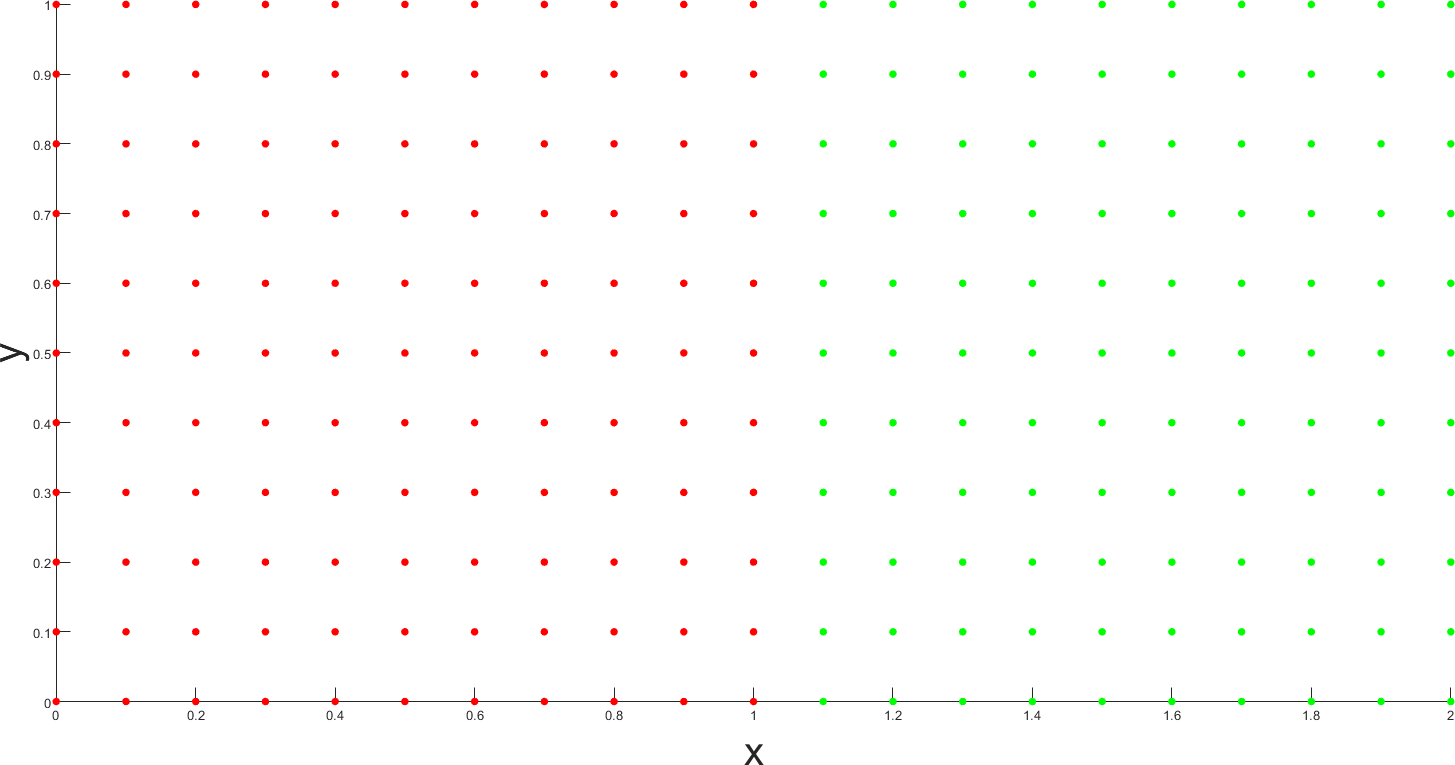}
    \caption{}
    \sublabel{fig:PoincareIC}
    \end{subfigure}
    \begin{subfigure}[t]{0.48\textwidth}
    \includegraphics[width=\textwidth]{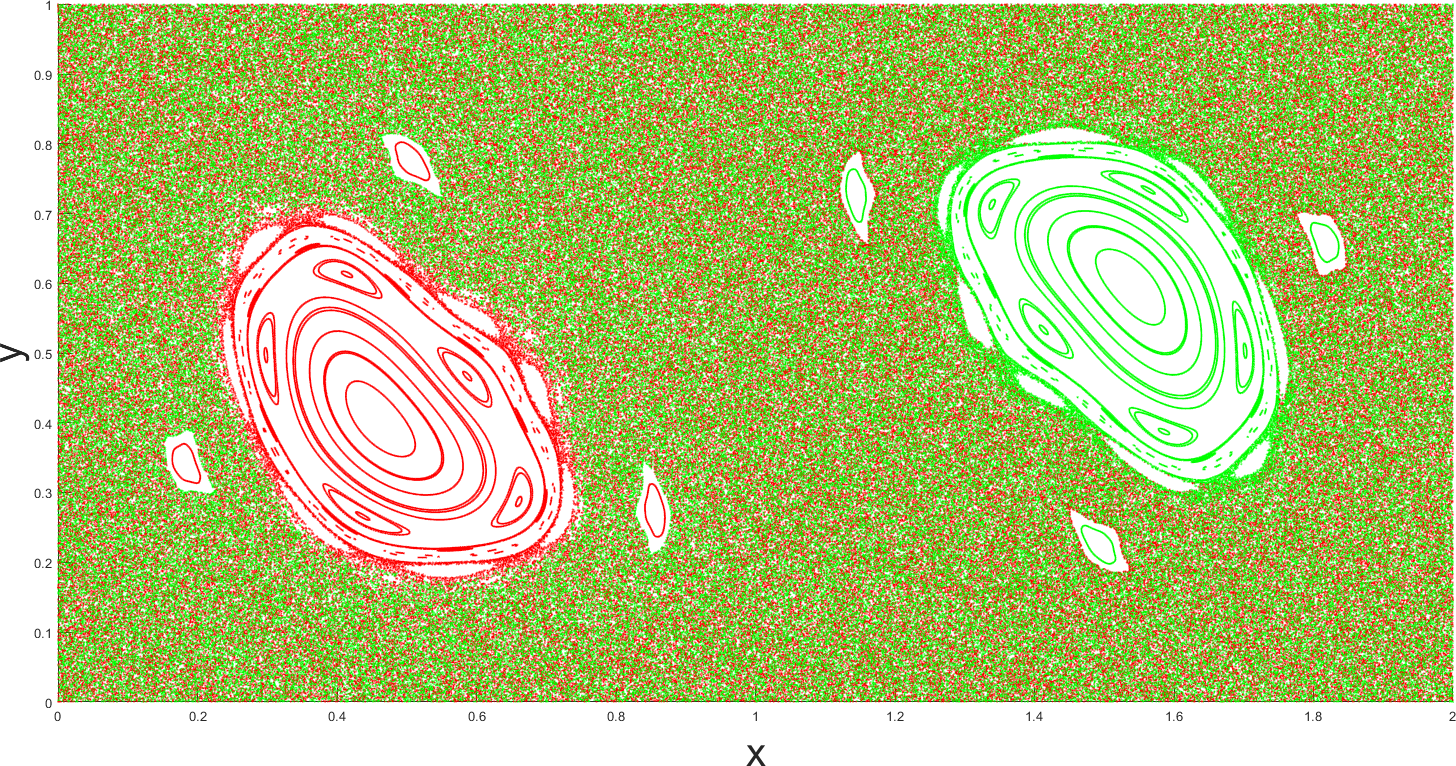}
    \caption{}
    \sublabel{fig:PoincareF}
    \end{subfigure}
\caption{Poincare map of the double-gyre such as in Equations \eqref{eqn:UPsi} and \eqref{eqn:timedep} with parameters $\epsilon = 0.25$ and $\omega = 2$. 4000 iterations of the period of perturbation are shown. Points with initial conditions to the left of $x = 1$ are shown in red and points with initial conditions to the right of $x = 1$ are shown in green. Invariant tori and destroyed tori can be seen in the map, as solid or broken curves respectively.}
\label{fig:Poincare}
\end{figure}

Our understanding of KAM theory and Poincare Maps now gives us one way to interpret the results from Figure \ref{fig:twotrajs}. Comparing Figures \ref{fig:twotrajs} and \ref{fig:PoincareF}, one sees that the trajectory which remained in the gyre was initialized inside of a KAM invariant tori, and so it was constrained within an ``island'' without the ability to escape. Similarly, the other trajectory in Figure \ref{fig:twotrajs} was initialized outside of the invariant tori, and as such, was able to move about the domain chaotically.  \\

\subsubsection{Relationship between LDs and KAM Tori}

\hspace{\parindent} In side-by-side examples, KAM invariant tori seem to appear in Lagrangian Descriptors as uniform ``dips.'' It has also been shown that there is a rigorous connection between invariant sets (such as KAM invariant tori) and the time-average of functions along trajectories (such as Lagrangian Descriptors). This connection is based on Birkhoff's ergodic theorem and has been discussed in \cite{mezic1999}, \cite{susuki2009}, and more specifically with respect to Lagrangian Descriptors in \cite{lopesino2017}. The theorem states that the limit of time-averaged functions along trajectories (as $\tau \rightarrow \infty$) exists, so long as the dynamical system preserves smooth measures and is defined on a compact set. Level curves of these limit functions are invariant sets. Lopesino et al. note that this theory has not been generalized for aperiodic time-dependent flows \cite{lopesino2017}. In this paper, we reiterate a simple heuristic argument which connects KAM tori to Lagrangian descriptors. Closed trajectories within KAM invariant tori will have trajectories of similar past and future behavior. Being ``trapped'' within their tori, these trajectories are also less likely to have seen much of the domain, leading to lower LD values. Outside of these tori are regions of higher complexity, that are associated with strong mixing (and consequently, trajectories that have ``visited'' more of the domain and are likely to have higher LD values). \\

\begin{figure}[H]
\begin{subfigure}[t]{0.33\textwidth}
\includegraphics[width=\textwidth]{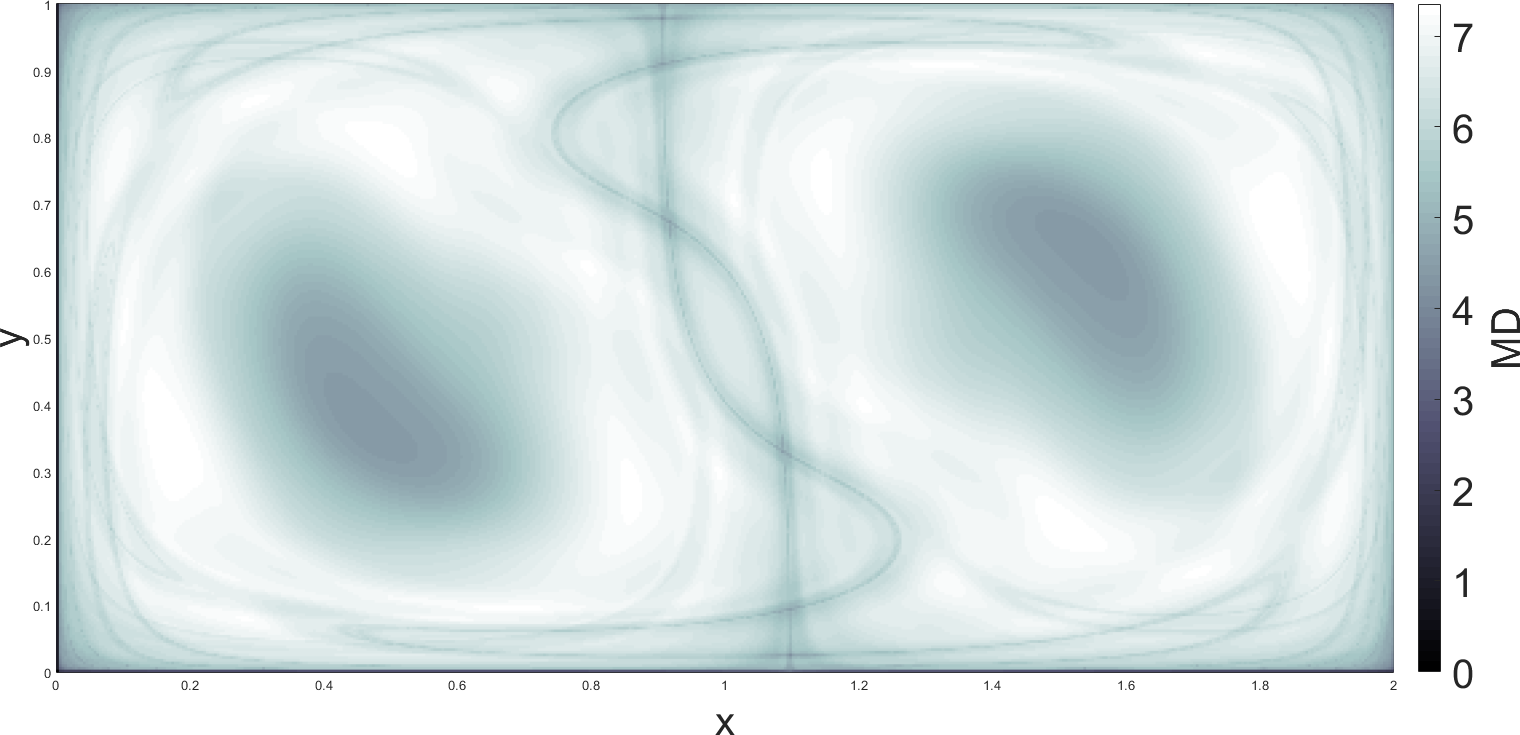}
\caption{}
\sublabel{fig:PoincareLD1}
\end{subfigure}\hfill
\begin{subfigure}[t]{0.31\textwidth}
\includegraphics[width=\textwidth]{Poincare_grid_epsilon25_omega_20_its4000.png}
\caption{}
\sublabel{fig:PoincareLD2}
\end{subfigure}\hfill
\centering
\begin{subfigure}[t]{0.33\textwidth}
\includegraphics[width=\textwidth]{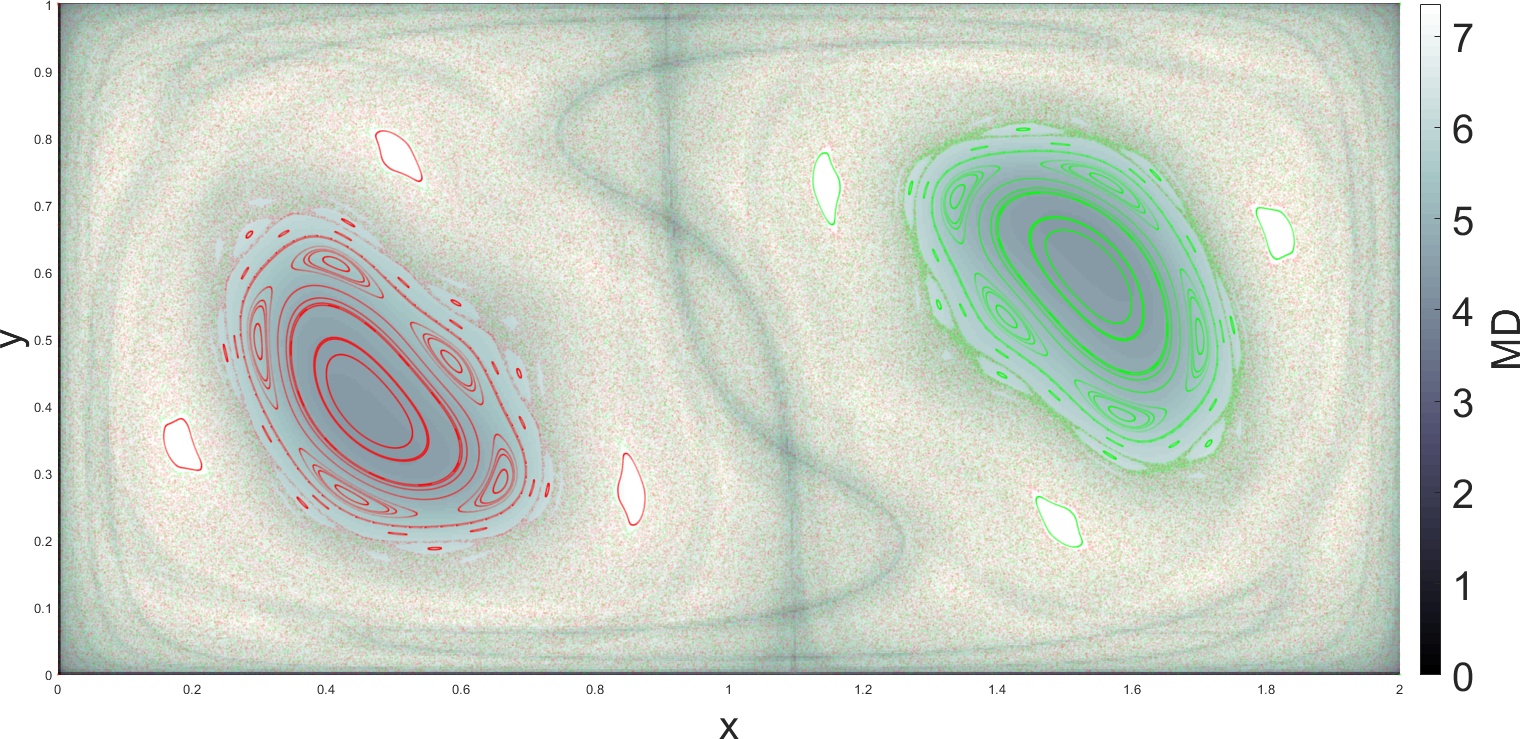}
\caption{}
\sublabel{fig:PoincareLD3}
\end{subfigure}
\caption{LDs and KAM tori in the double-gyre such as in Equations \eqref{eqn:UPsi} and \eqref{eqn:timedep} with parameters $A = .1$, $\epsilon = .25$ and $\omega = 2$. Panel (a) shows an LD field using $\tau = 15$, panel (b) shows a Poincare map up to 4000 iterations (identical to Figure \ref{fig:Poincare}), and panel (c) has shown the Poincare map overlaid onto the LD field.}
\label{fig:Poincare3panel}
\end{figure}

Figure \ref{fig:Poincare3panel} contains three panels: \ref{fig:PoincareLD1} shows an LD field for the double-gyre with parameters $A = .1$, $\epsilon = 0.25$, $\omega = 2$, and $\tau = 15$, \ref{fig:PoincareLD2} shows a Poincare map for the same system representing 4000 iterations, and \ref{fig:PoincareLD3} shows the Poincare map from (b) overlaid onto the LD field in (a). One can see that the KAM tori correspond nicely with the boundaries of the LD ``islands.'' More examples of this can be seen throughout the results in Section \ref{sec:results}. 

\section{Results} \label{sec:results}

\hspace{\parindent} Now that both FTLEs and LDs have been discussed in depth, we will move on to a series of results which compare the two diagnostics using varying parameters of the time-dependent double-gyre. Figures 6-14 contain examples of FTLEs, LDs, and Poincare maps for various parameters in the time-dependent double-gyre. The parameters $A = .1$ and $t_0 = 0$ are used for all cases. The FTLE panels contain the sum of the forward and backward FTLE fields for each case. FTLEs and LDs are computed for $\tau = 15$, and Poincare maps are shown for 4000 iterations of the period of perturbation. The set of initial conditions consist of a $501\times251$ grid for FTLEs and LDs, and a $21\times11$ grid for Poincare Maps. For Poincare Maps, points are shown in red if their respective initial condition began on the left side of the domain ($x < 1$) or green if they begin on the right side of the domain ($x > 1$). Computing trajectories was done using the velocity field described by Equations \eqref{eqn:UPsi} and \eqref{eqn:timedep}, while using \textit{MATLAB}'s ode45 function, which implements an explicit 4th-order variable time step Runge-Kutta scheme.\\

We observe in the results that ridges of FTLE fields visually correspond to features in LD fields, except towards the ends of FTLE ridges, where the curves often become displaced or distorted. Sharp features in the FTLE and LD fields exist in the ``mixed'' chaotic regions of the Poincare maps (recall that the intersection of stable and unstable manifolds become a mechanism for transport between gyres).  Further, large Poincare ``islands'' are visibly related to smooth dips in LD fields and lack of ridges in the FTLE fields. 

\begin{figure}[H]
\begin{subfigure}[t]{0.33\textwidth}
\includegraphics[width=\textwidth]{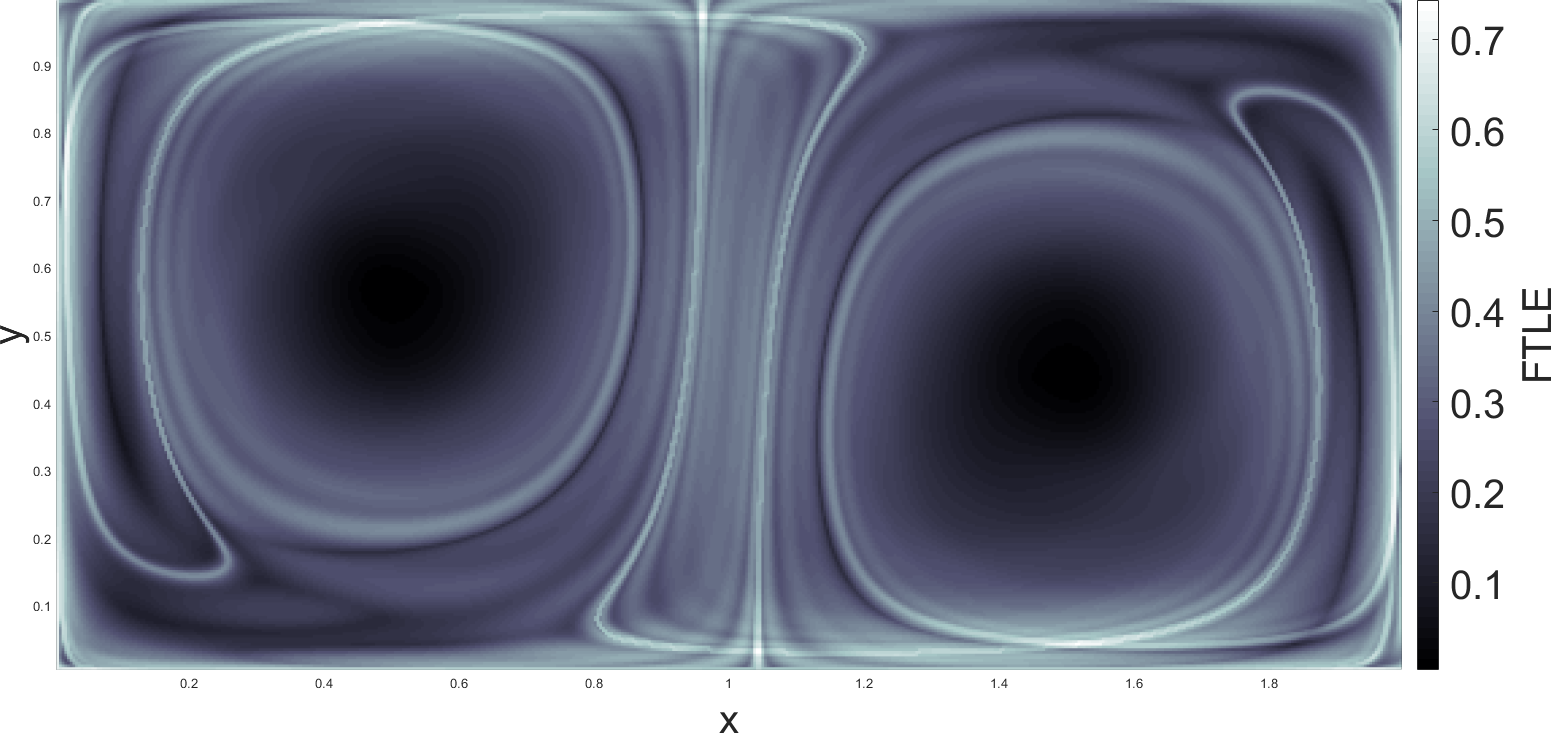}
\caption{}
\end{subfigure}\hfill
\begin{subfigure}[t]{0.33\textwidth}
\includegraphics[width=\textwidth]{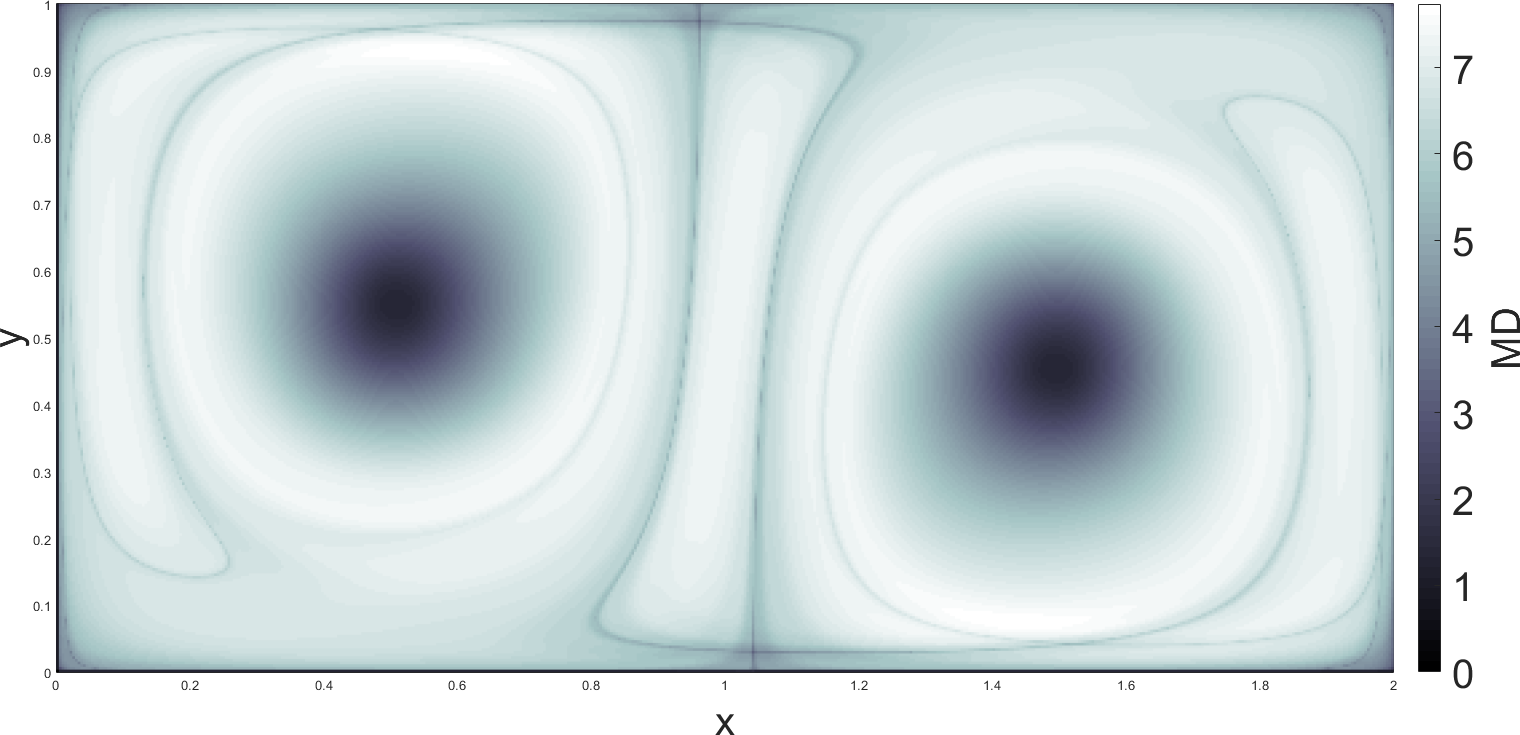}
\caption{}
\end{subfigure}\hfill
\begin{subfigure}[t]{0.31\textwidth}
\includegraphics[width=\textwidth]{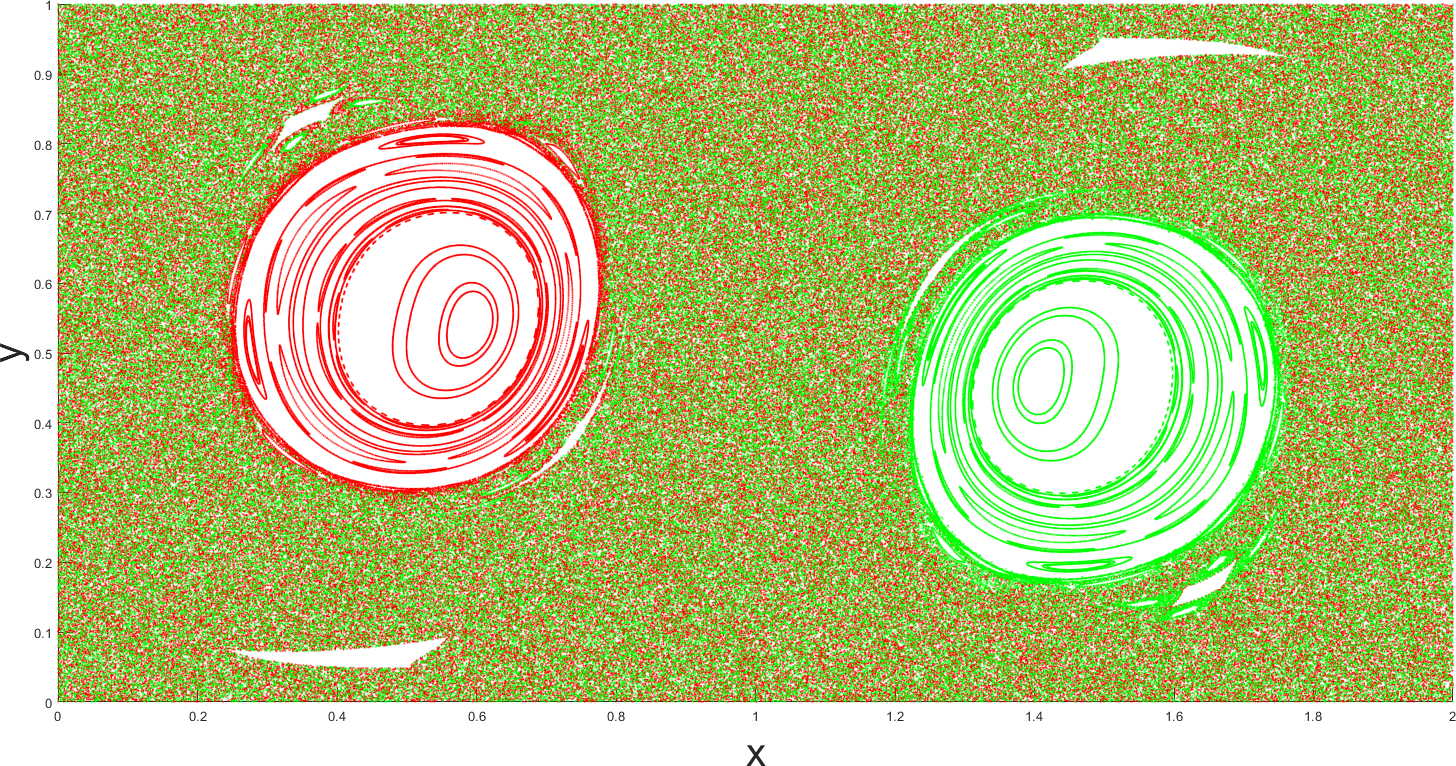}
\caption{}
\end{subfigure}
\begin{subfigure}[t]{0.33\textwidth}
\includegraphics[width=\textwidth]{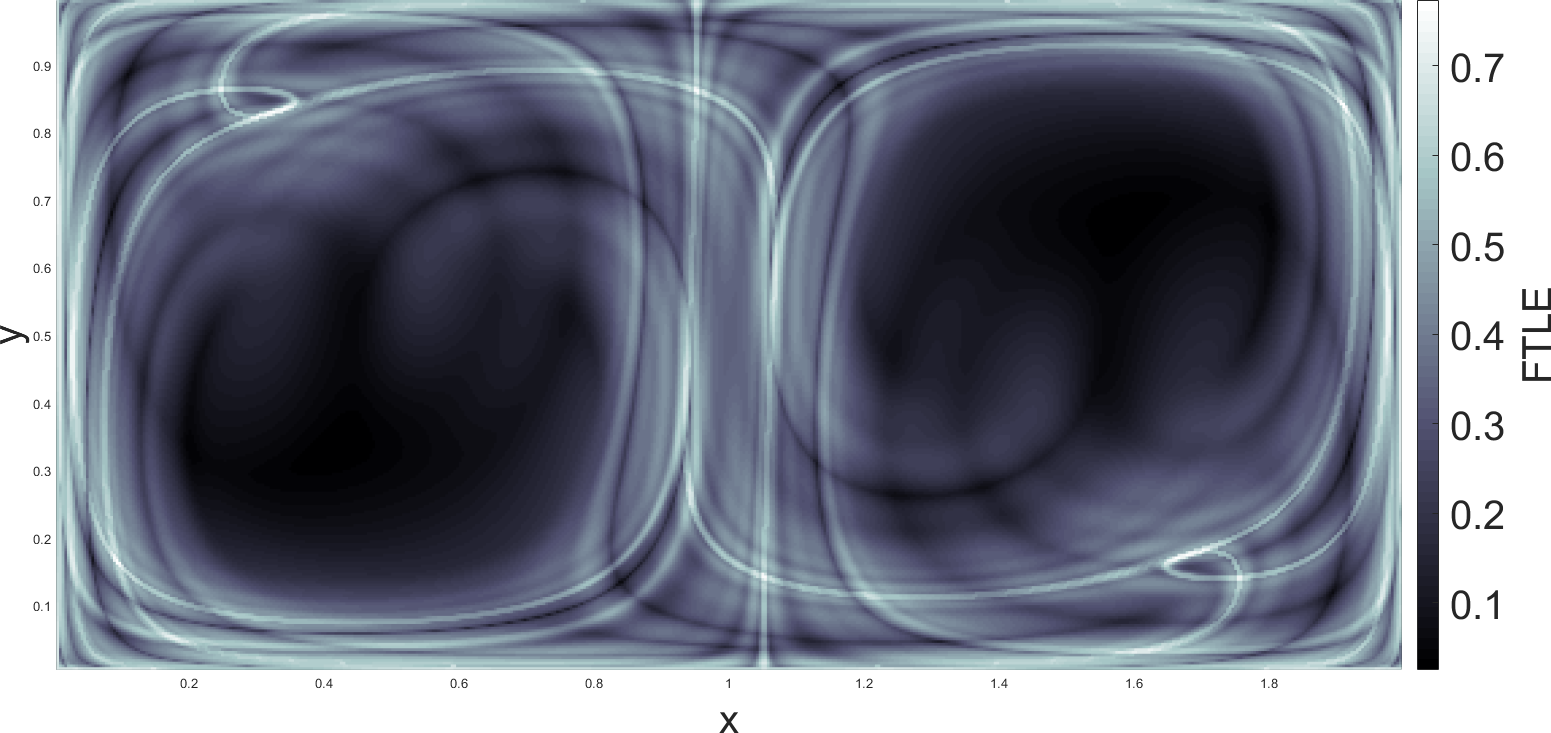}
\caption{}
\end{subfigure}\hfill
\begin{subfigure}[t]{0.33\textwidth}
\includegraphics[width=\textwidth]{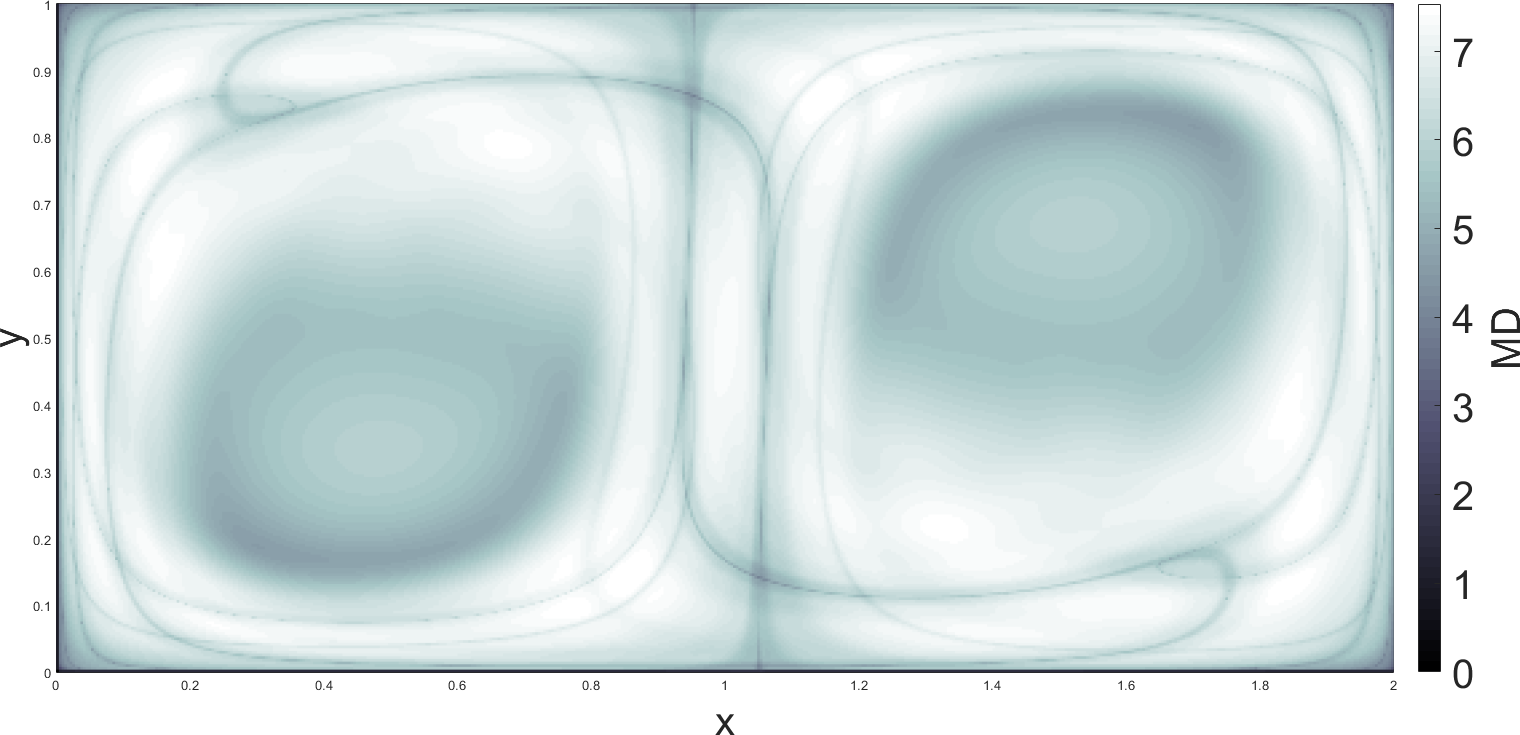}
\caption{}
\end{subfigure}\hfill
\begin{subfigure}[t]{0.31\textwidth}
\includegraphics[width=\textwidth]{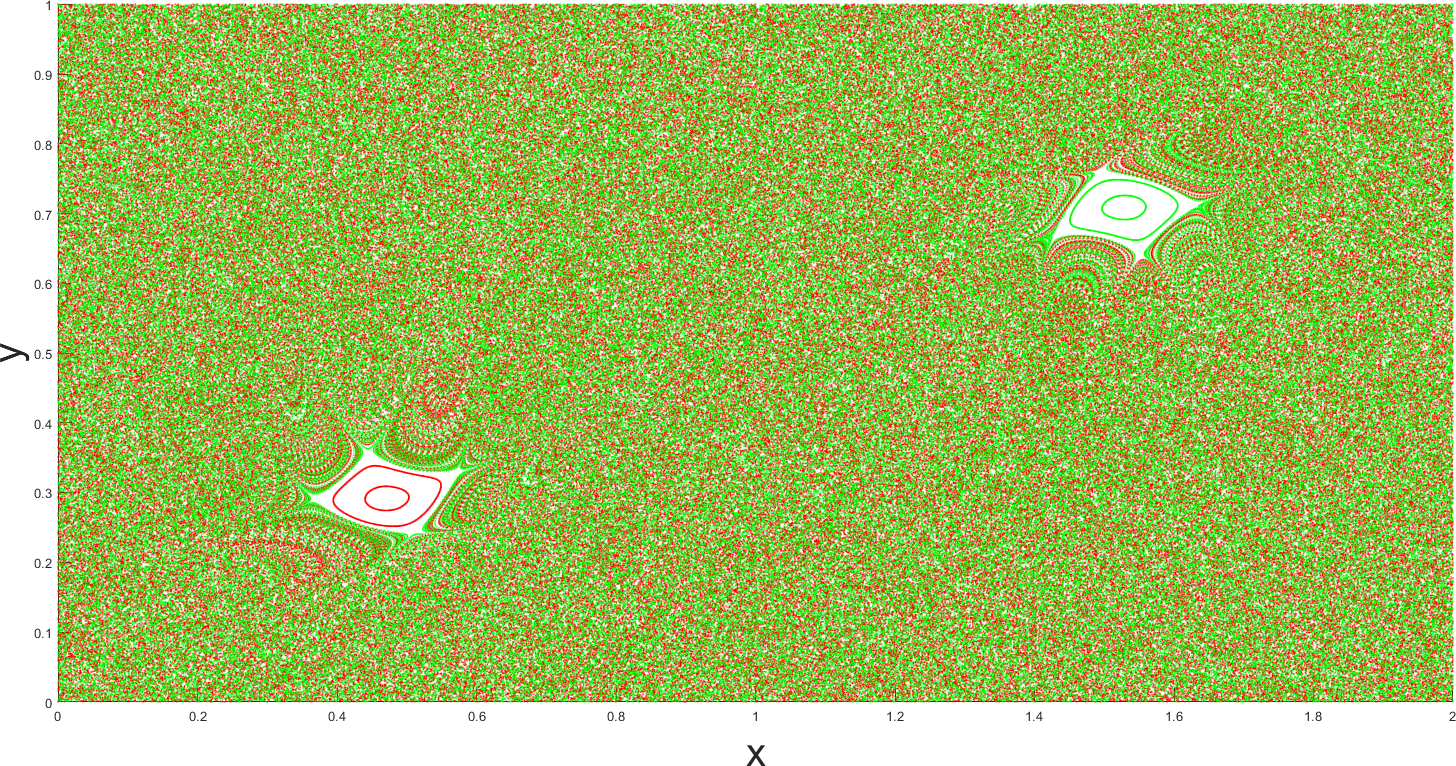}
\caption{}
\end{subfigure}
\begin{subfigure}[t]{0.33\textwidth}
\includegraphics[width=\textwidth]{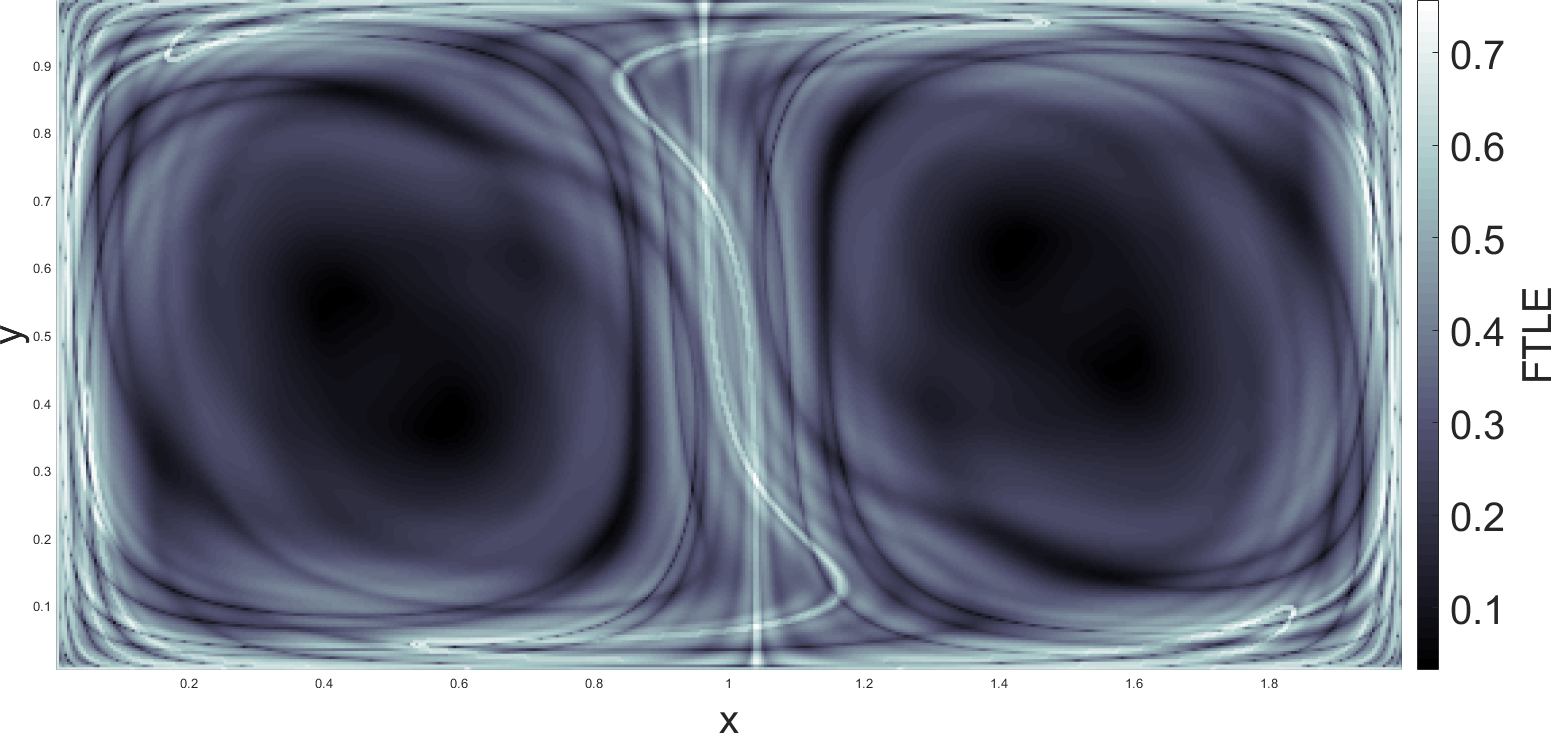}
\caption{}
\end{subfigure}\hfill
\begin{subfigure}[t]{0.33\textwidth}
\includegraphics[width=\textwidth]{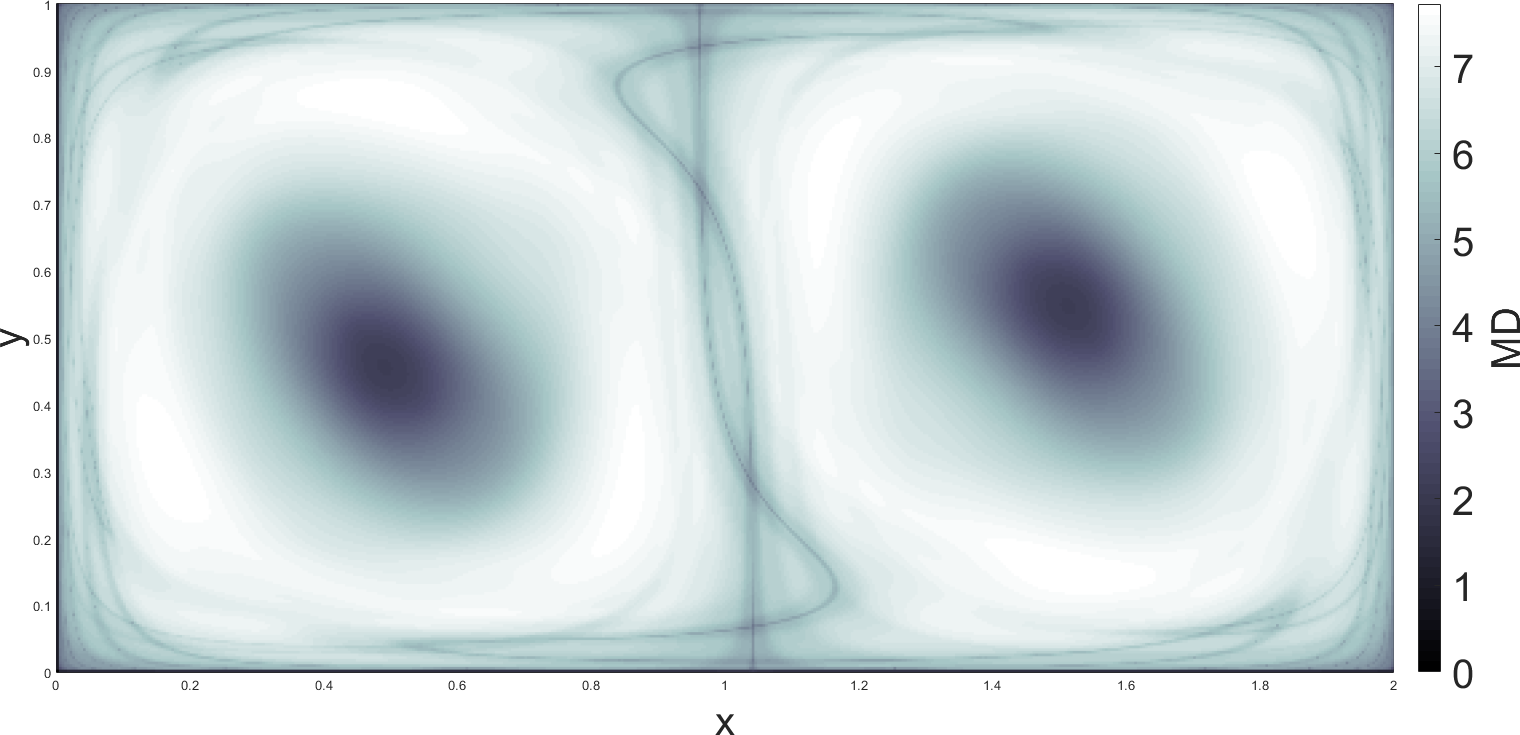}
\caption{}
\end{subfigure}\hfill
\begin{subfigure}[t]{0.31\textwidth}
\includegraphics[width=\textwidth]{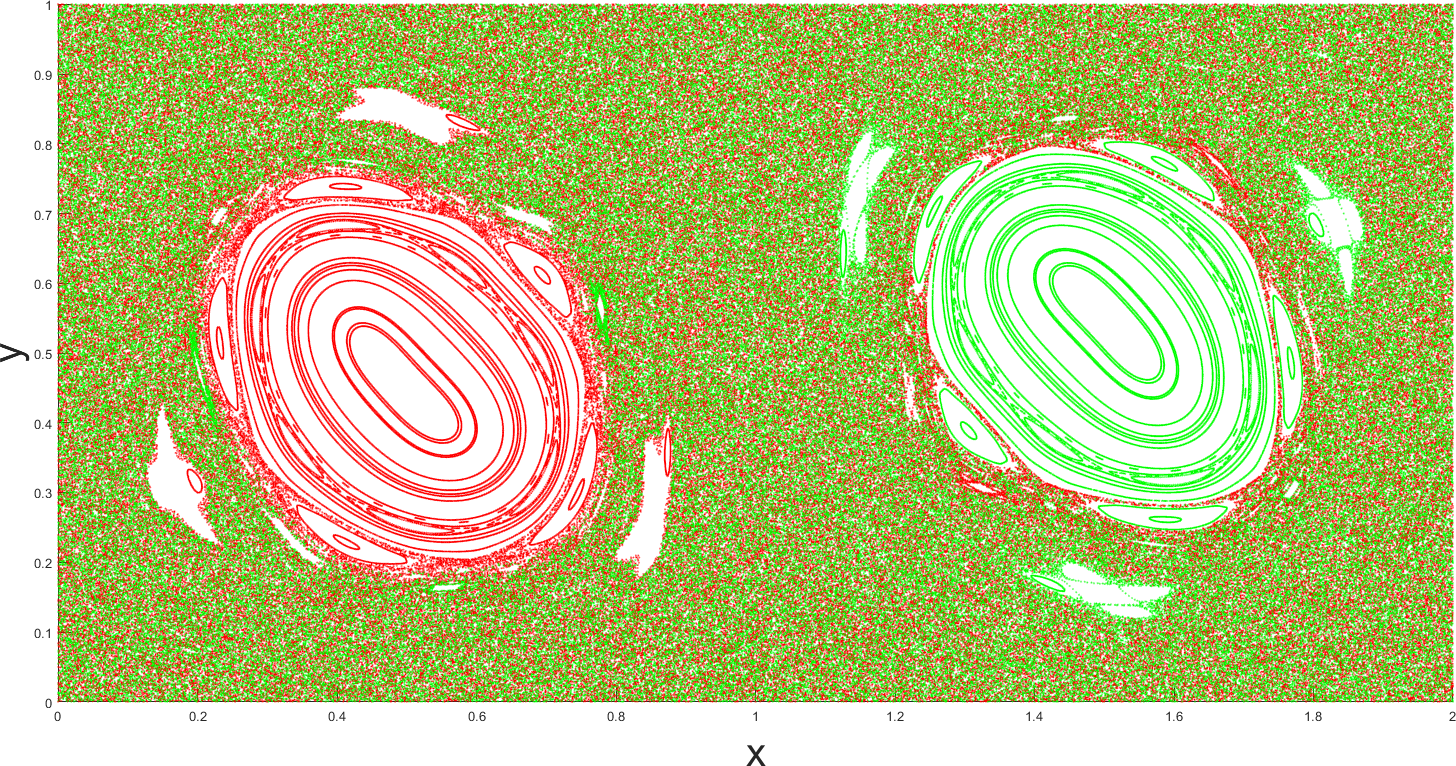}
\caption{}
\end{subfigure}
\caption{LDs, FTLEs, and KAM tori in the double-gyre such as in Equations \eqref{eqn:UPsi} and \eqref{eqn:timedep} with parameters $A = .1$ and $\epsilon = .1$. The first row [(a),(b),(c)] corresponds to $\omega = .5$, the second row [(d),(e),(f)] corresponds to $\omega = 1$, and the third row [(g),(h),(i)] corresponds to $\omega = 2$. The first column [(a),(d),(g)] contains summed forward and backward FTLE fields with $\tau = 15$, the second column [(b),(e),(h)] contains LD fields with $\tau = 15$, and the third column [(c),(f),(i)] contains Poincare Maps up to 4000 iterations of the period of perturbation.}
\end{figure}

\begin{figure}[H]
\begin{subfigure}[t]{0.33\textwidth}
\includegraphics[width=\textwidth]{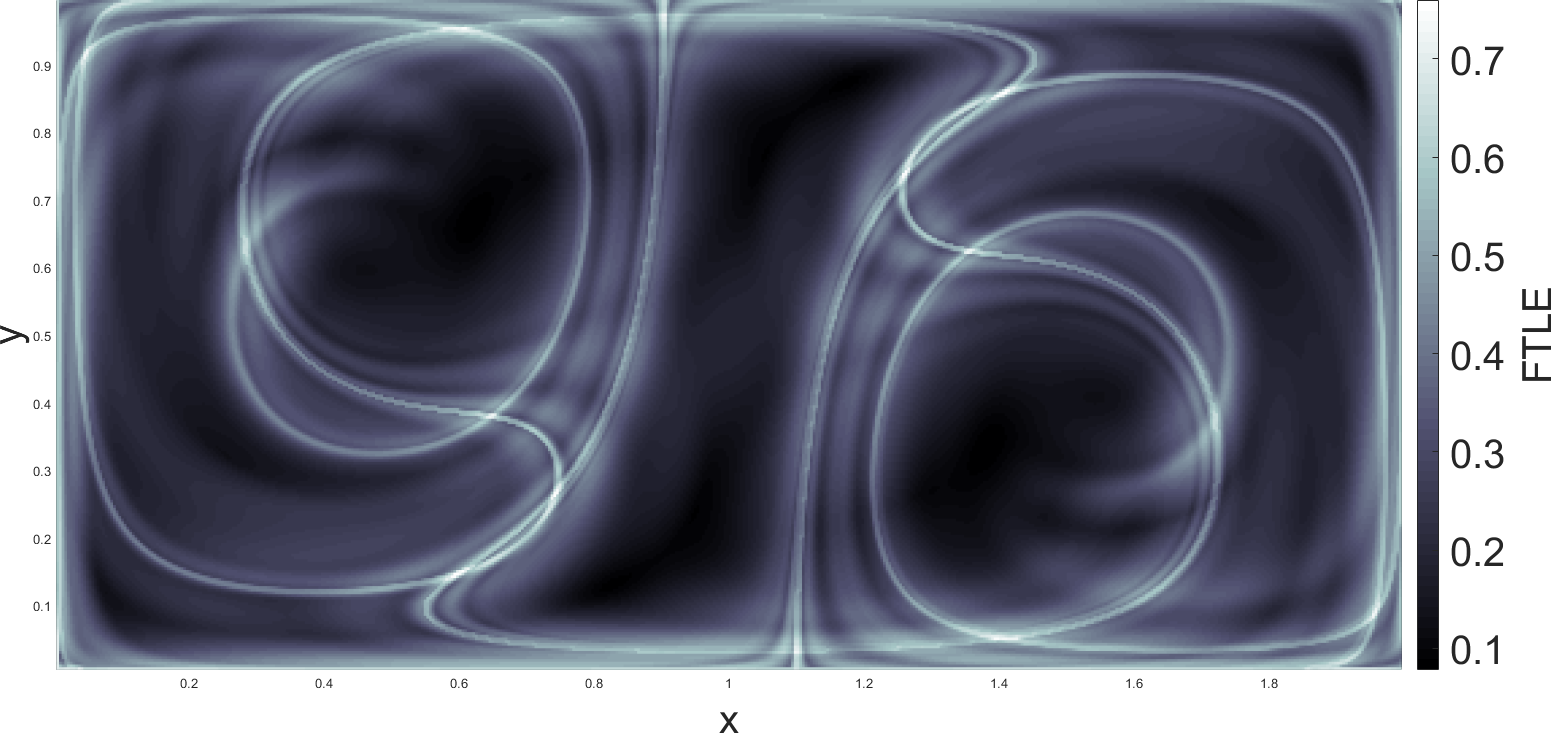}
\caption{}
\end{subfigure}\hfill
\begin{subfigure}[t]{0.33\textwidth}
\includegraphics[width=\textwidth]{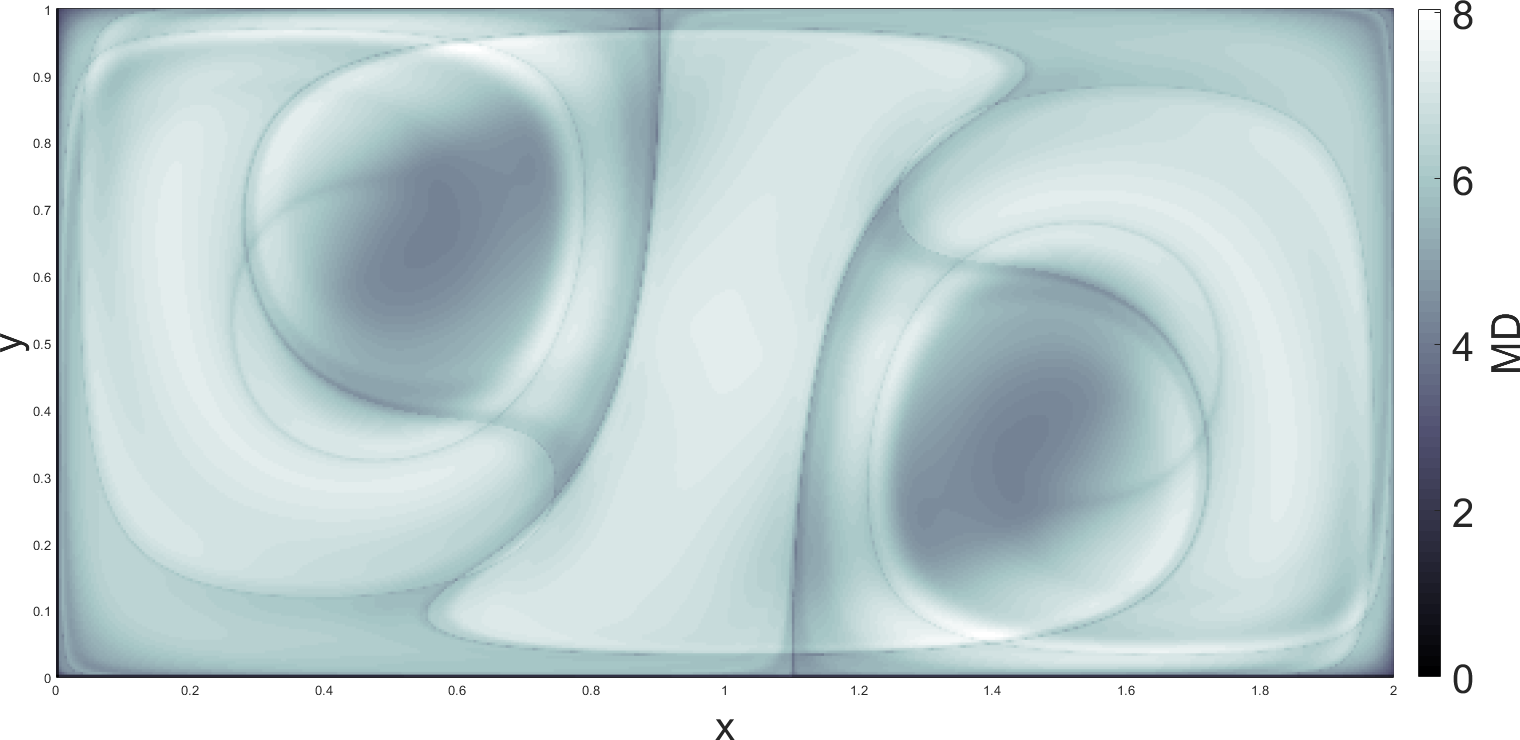}
\caption{}
\end{subfigure}\hfill
\begin{subfigure}[t]{0.31\textwidth}
\includegraphics[width=\textwidth]{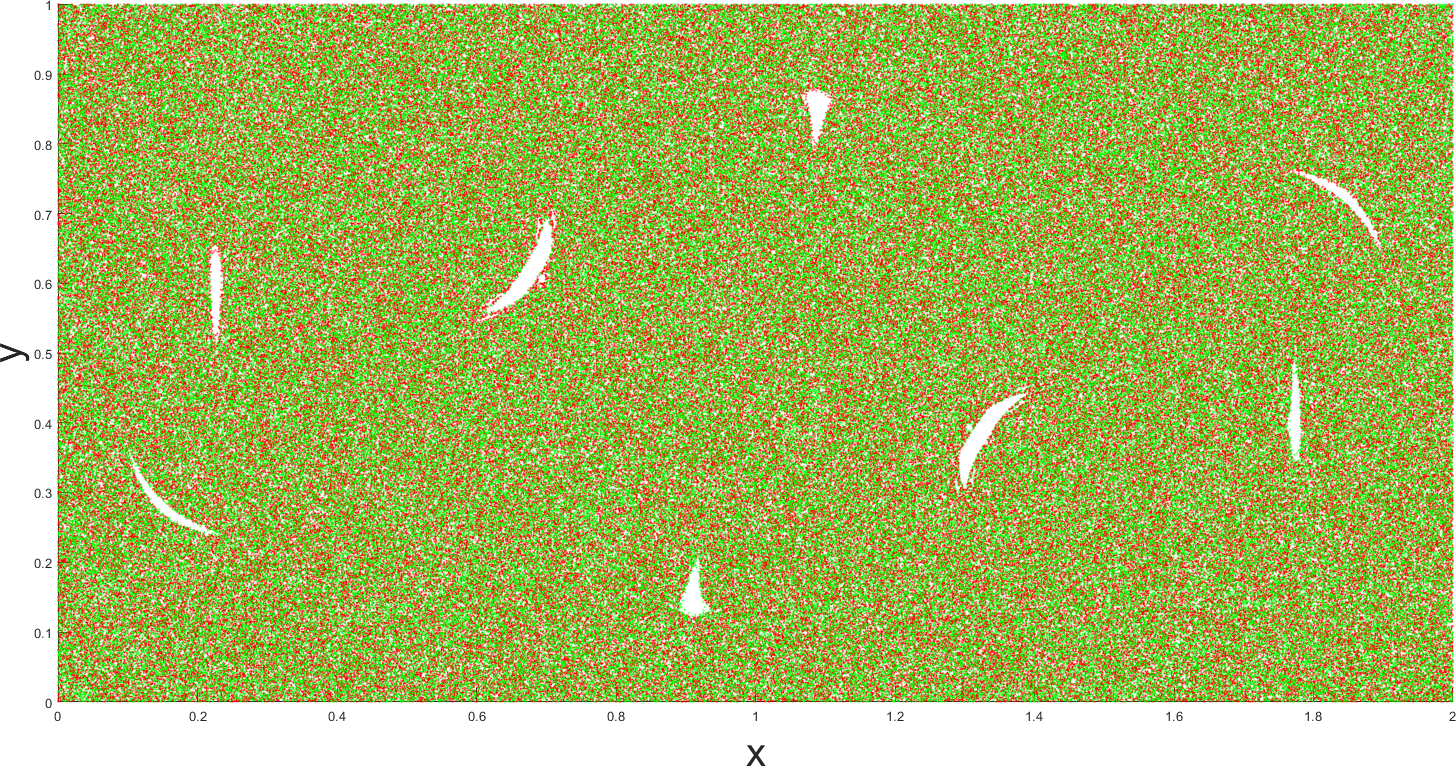}
\caption{}
\end{subfigure}
\begin{subfigure}[t]{0.33\textwidth}
\includegraphics[width=\textwidth]{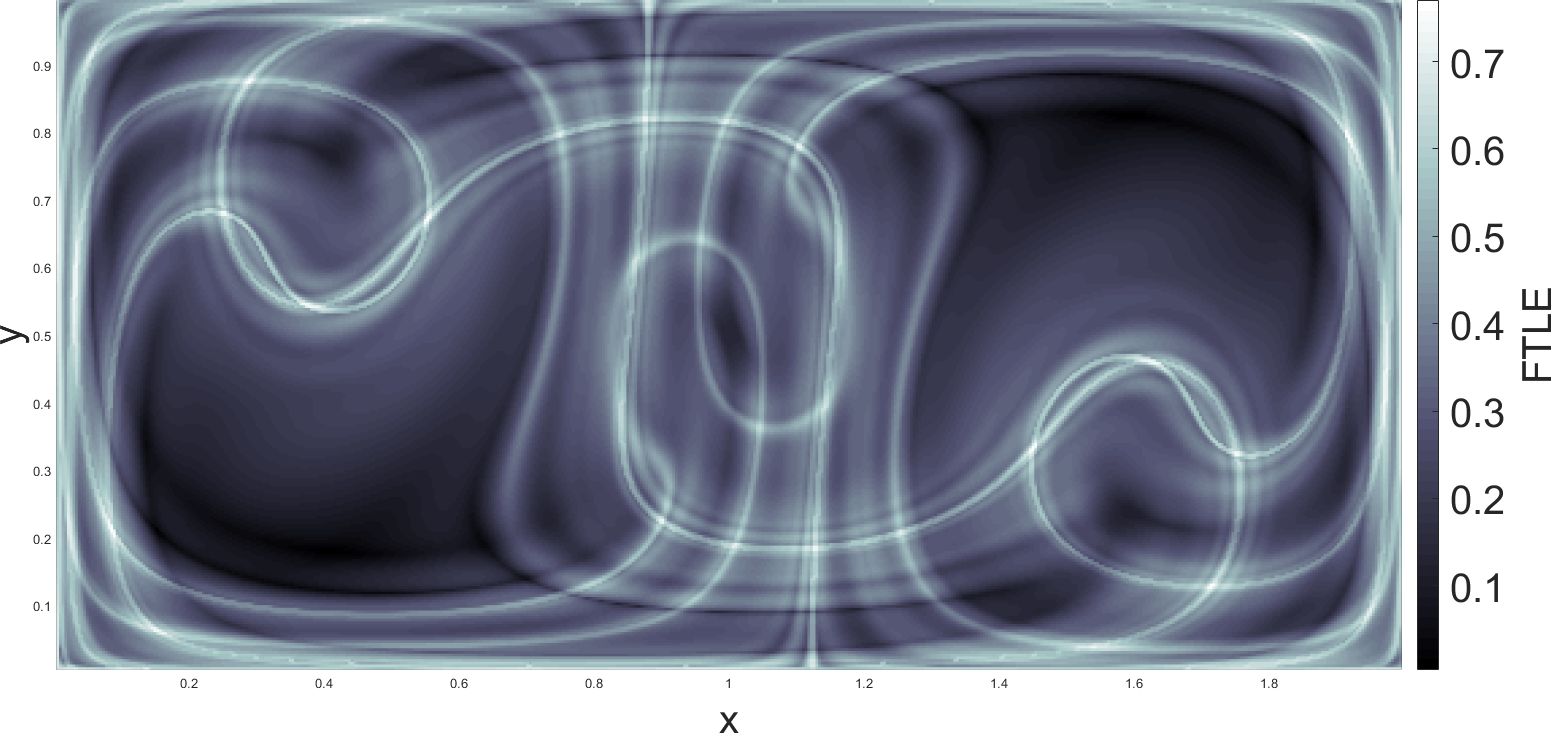}
\caption{}
\end{subfigure}\hfill
\begin{subfigure}[t]{0.33\textwidth}
\includegraphics[width=\textwidth]{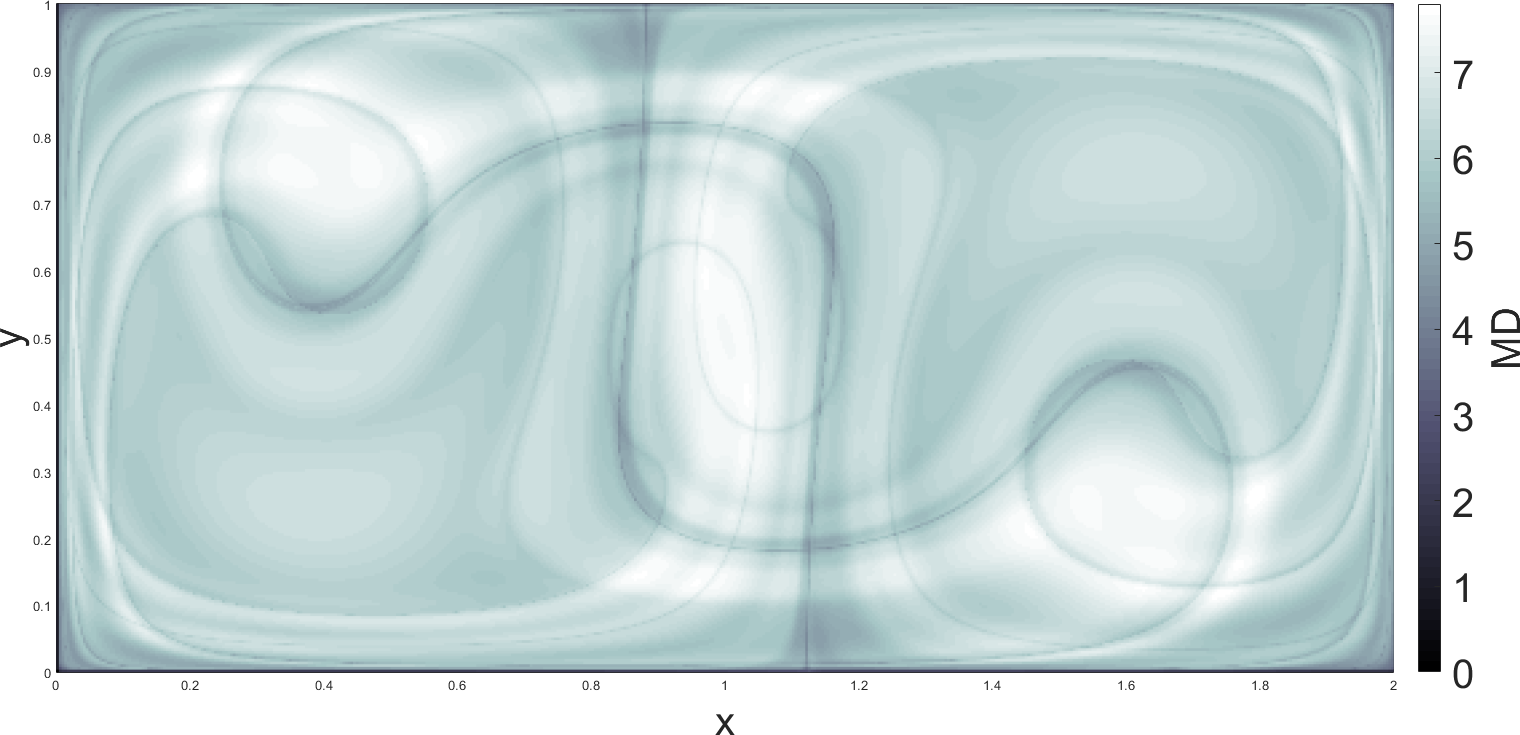}
\caption{}
\end{subfigure}\hfill
\begin{subfigure}[t]{0.31\textwidth}
\includegraphics[width=\textwidth]{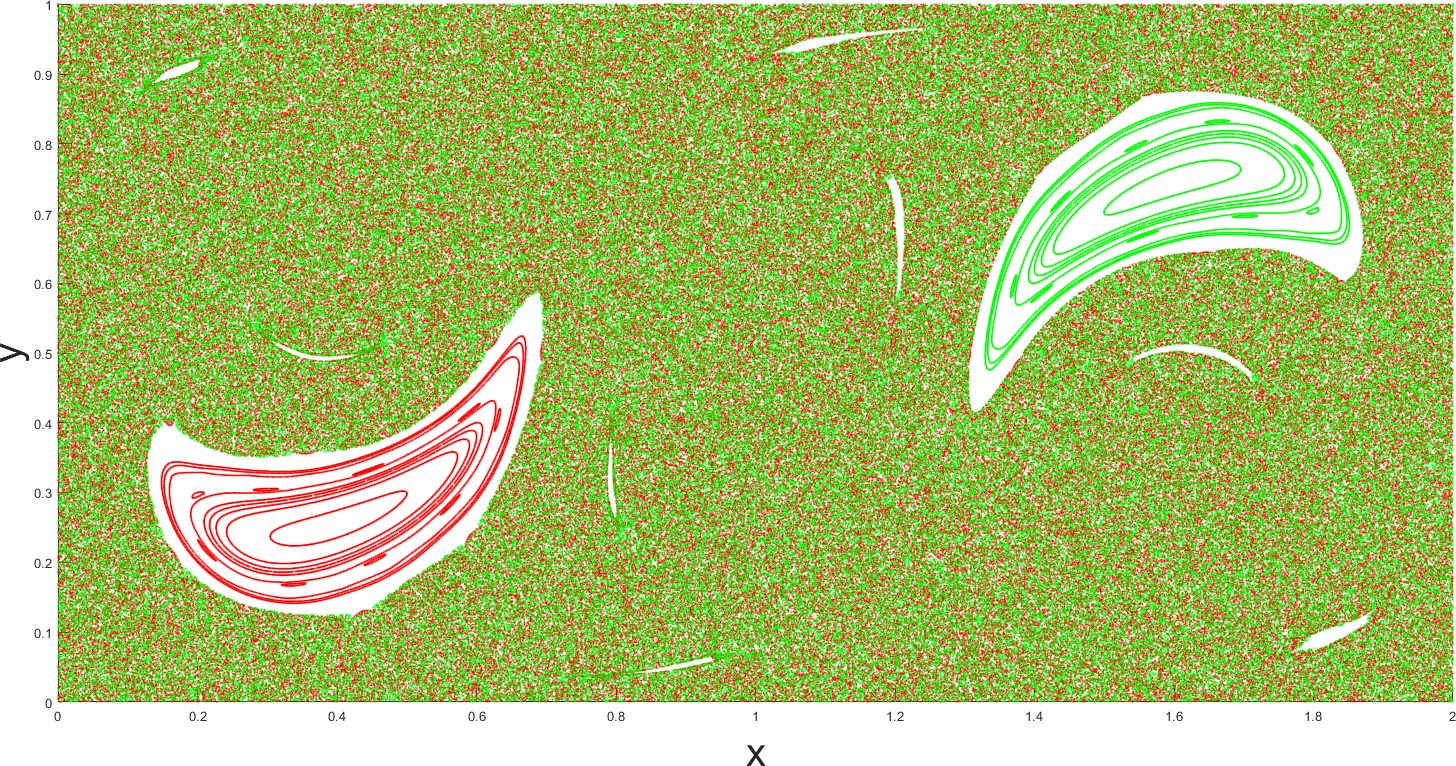}
\caption{}
\end{subfigure}
\begin{subfigure}[t]{0.33\textwidth}
\includegraphics[width=\textwidth]{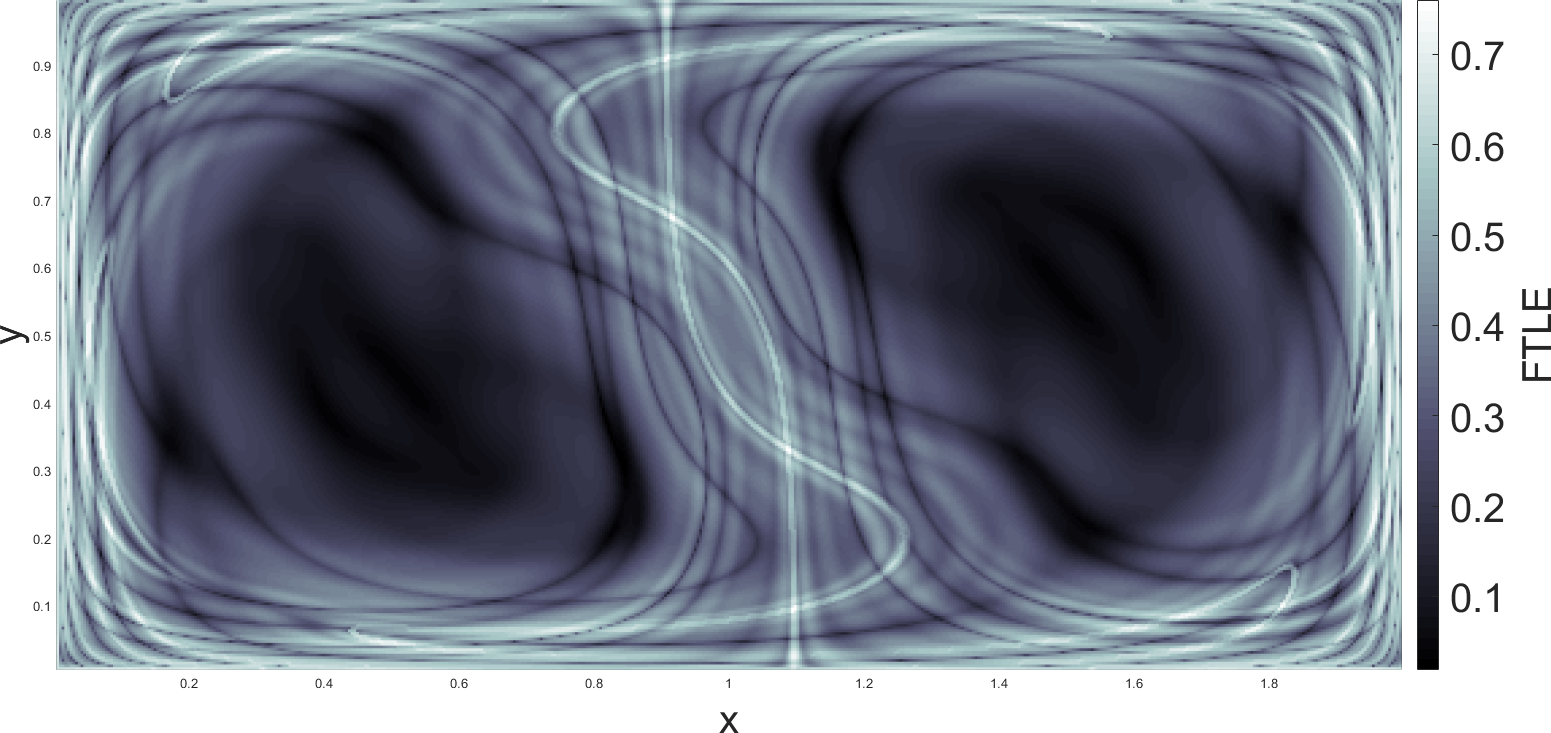}
\caption{}
\end{subfigure}\hfill
\begin{subfigure}[t]{0.33\textwidth}
\includegraphics[width=\textwidth]{LD_epsilon25_omega_20_tau15.png}
\caption{}
\end{subfigure}\hfill
\begin{subfigure}[t]{0.31\textwidth}
\includegraphics[width=\textwidth]{Poincare_grid_epsilon25_omega_20_its4000.png}
\caption{}
\end{subfigure}
\caption{LDs, FTLEs, and KAM tori in the double-gyre such as in Equations \eqref{eqn:UPsi} and \eqref{eqn:timedep} with parameters $A = .1$ and $\epsilon = .25$. The first row [(a),(b),(c)] corresponds to $\omega = .5$, the second row [(d),(e),(f)] corresponds to $\omega = 1$, and the third row [(g),(h),(i)] corresponds to $\omega = 2$. The first column [(a),(d),(g)] contains summed forward and backward FTLE fields with $\tau = 15$, the second column [(b),(e),(h)] contains LD fields with $\tau = 15$, and the third column [(c),(f),(i)] contains Poincare Maps up to 4000 iterations of the period of perturbation.}
\end{figure}

\begin{figure}[H]
\begin{subfigure}[t]{0.33\textwidth}
\includegraphics[width=\textwidth]{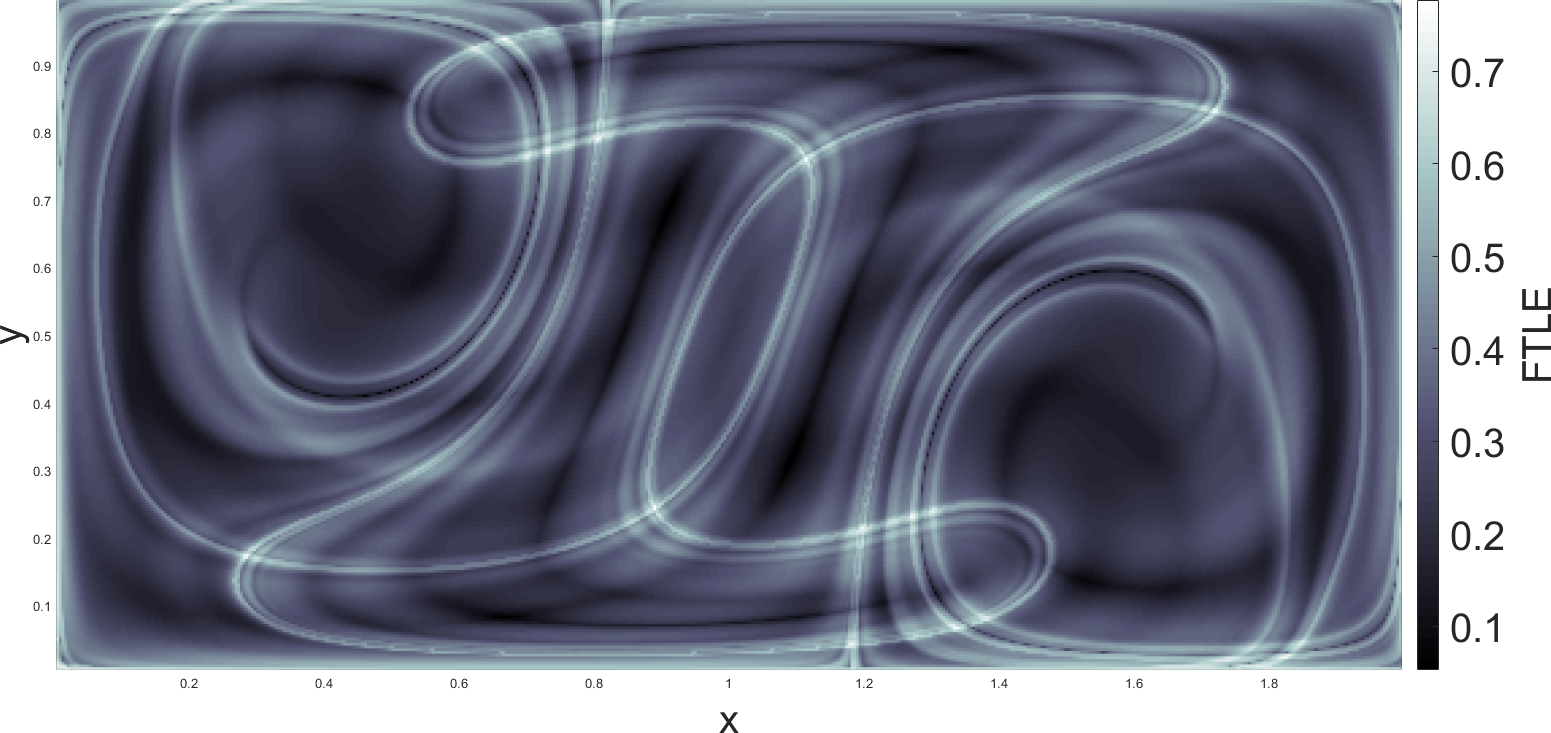}
\caption{}
\end{subfigure}\hfill
\begin{subfigure}[t]{0.33\textwidth}
\includegraphics[width=\textwidth]{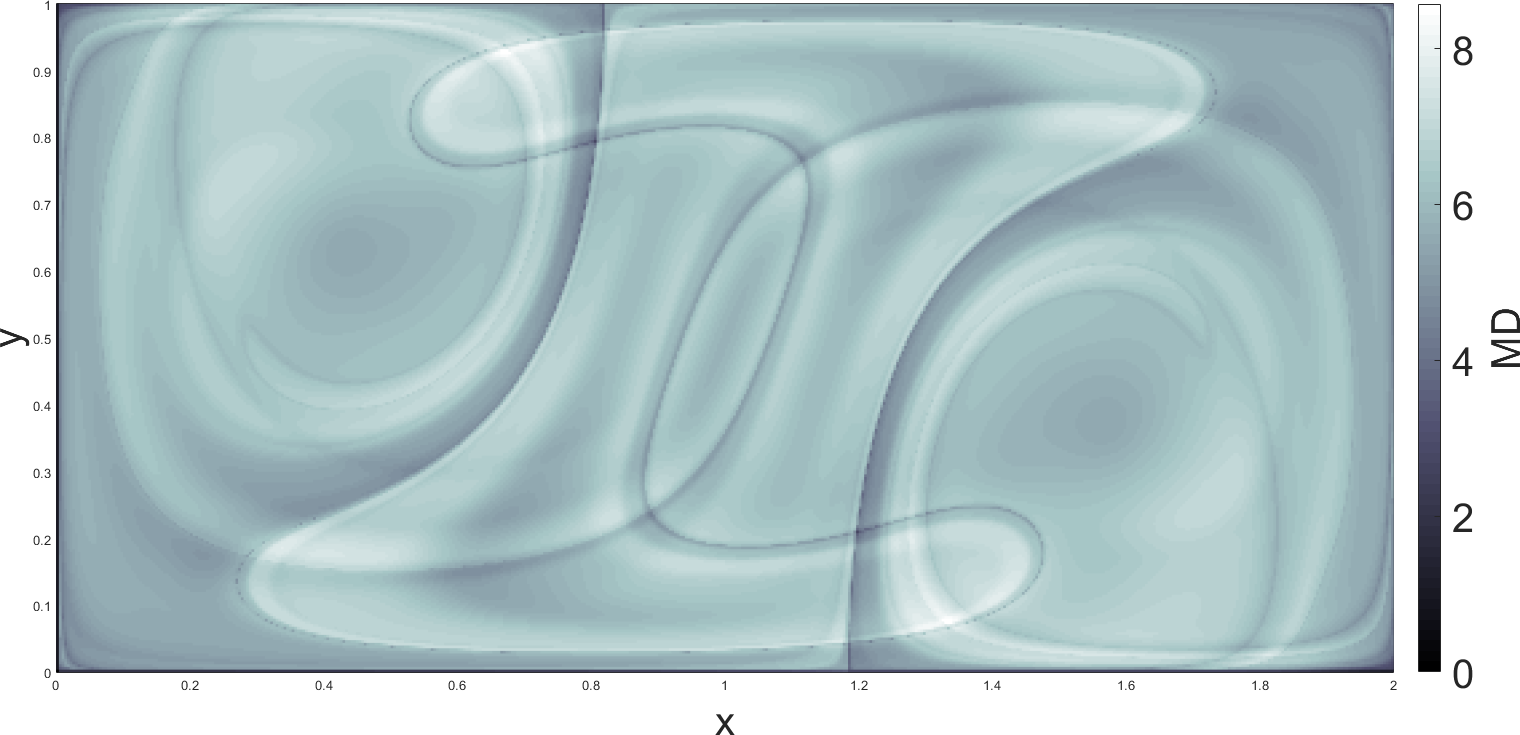}
\caption{}
\end{subfigure}\hfill
\begin{subfigure}[t]{0.31\textwidth}
\includegraphics[width=\textwidth]{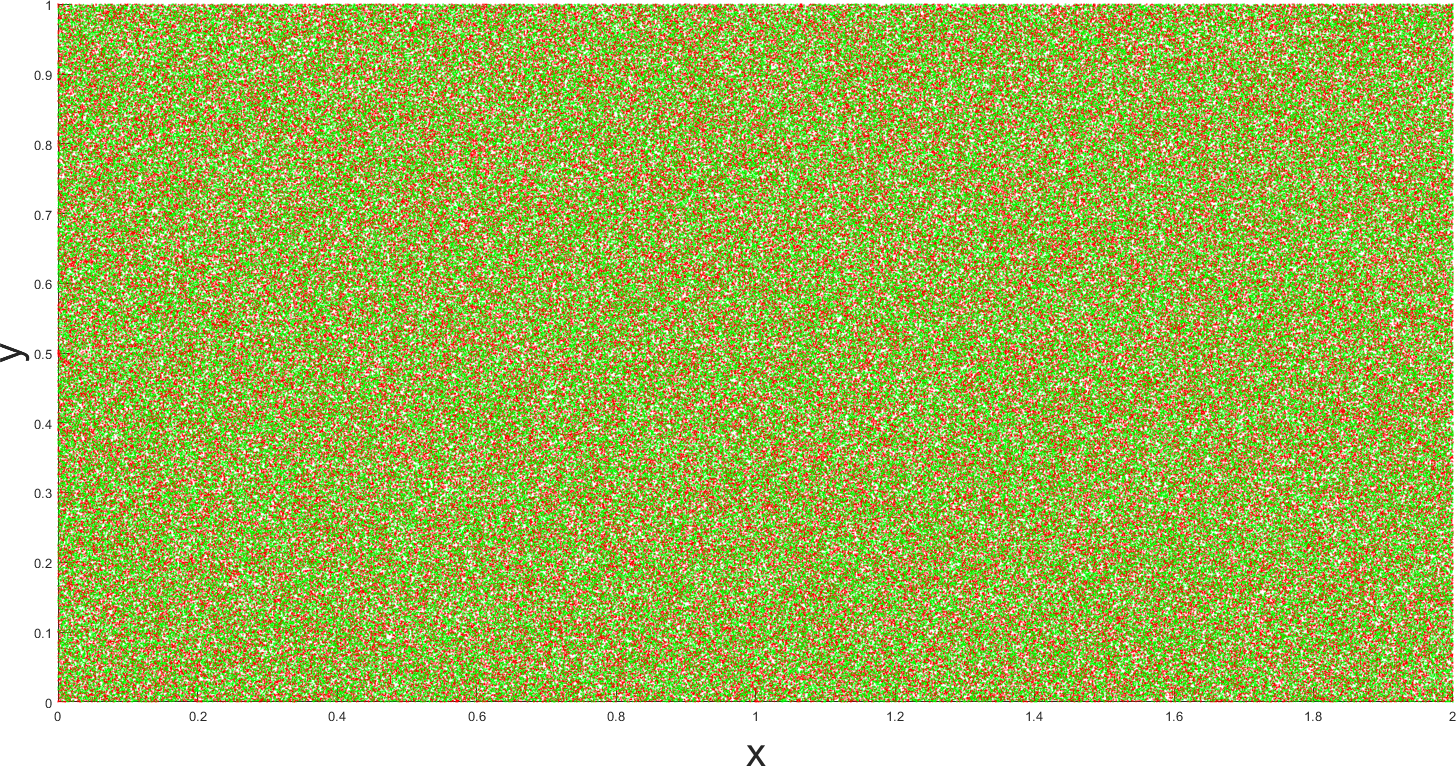}
\caption{}
\end{subfigure}
\begin{subfigure}[t]{0.33\textwidth}
\includegraphics[width=\textwidth]{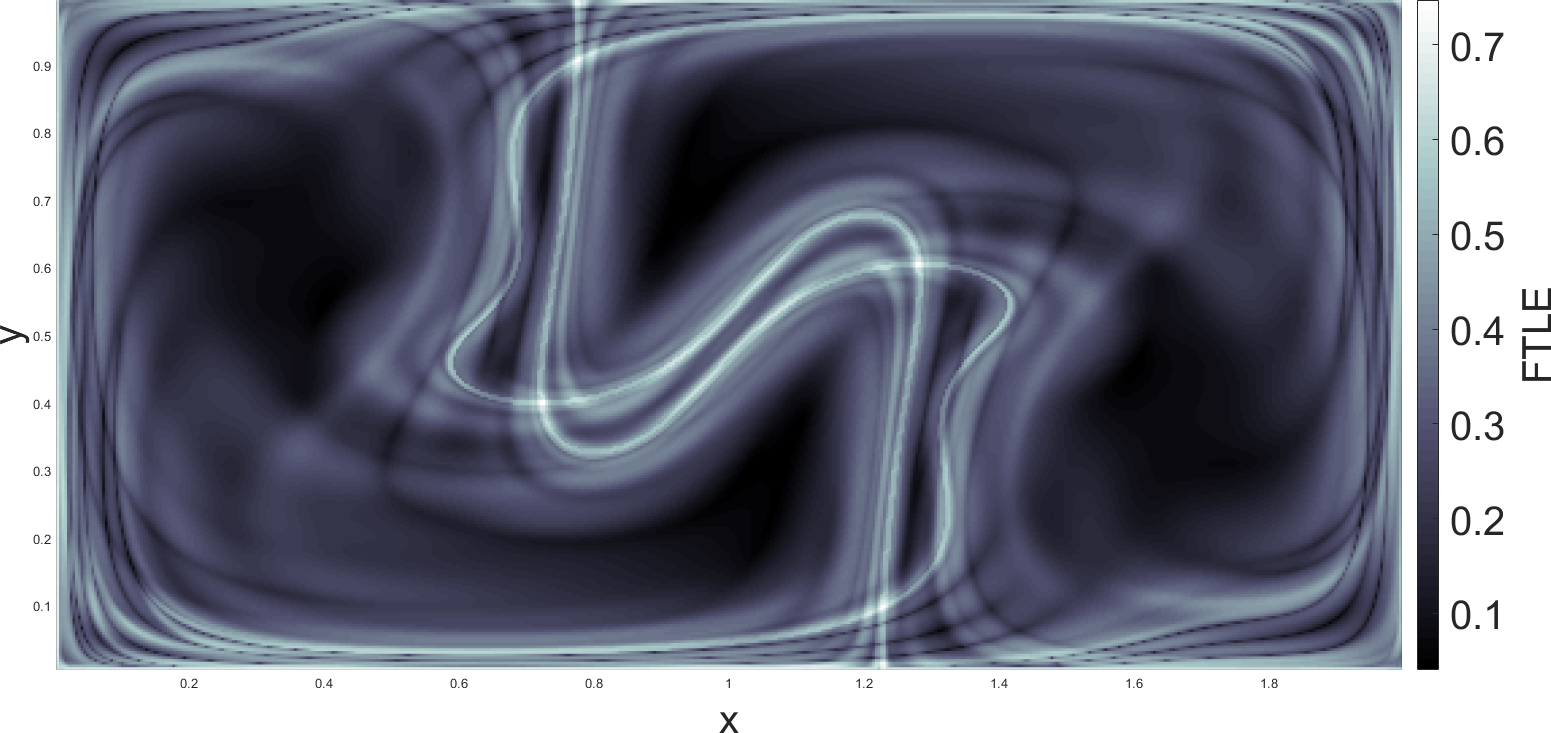}
\caption{}
\end{subfigure}\hfill
\begin{subfigure}[t]{0.33\textwidth}
\includegraphics[width=\textwidth]{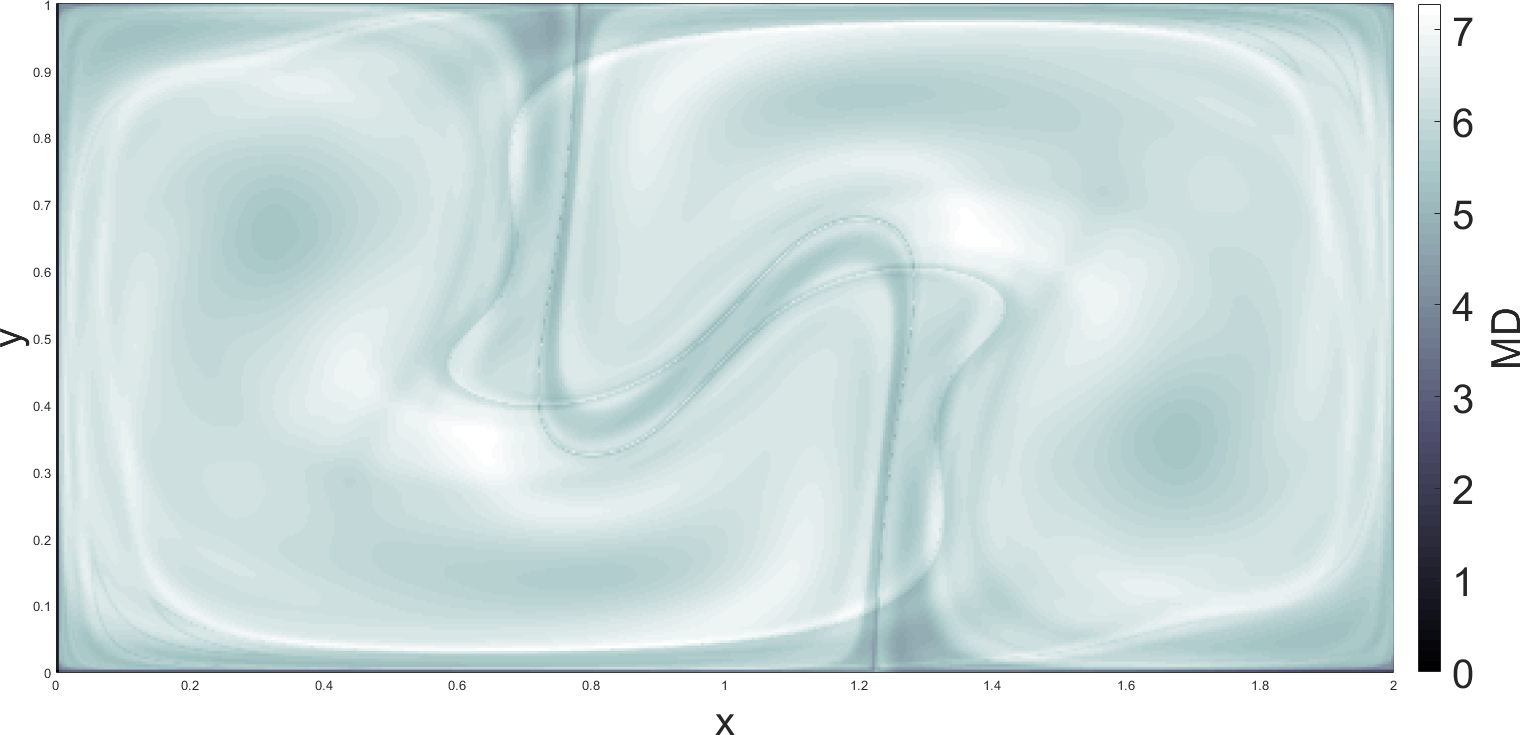}
\caption{}
\end{subfigure}\hfill
\begin{subfigure}[t]{0.31\textwidth}
\includegraphics[width=\textwidth]{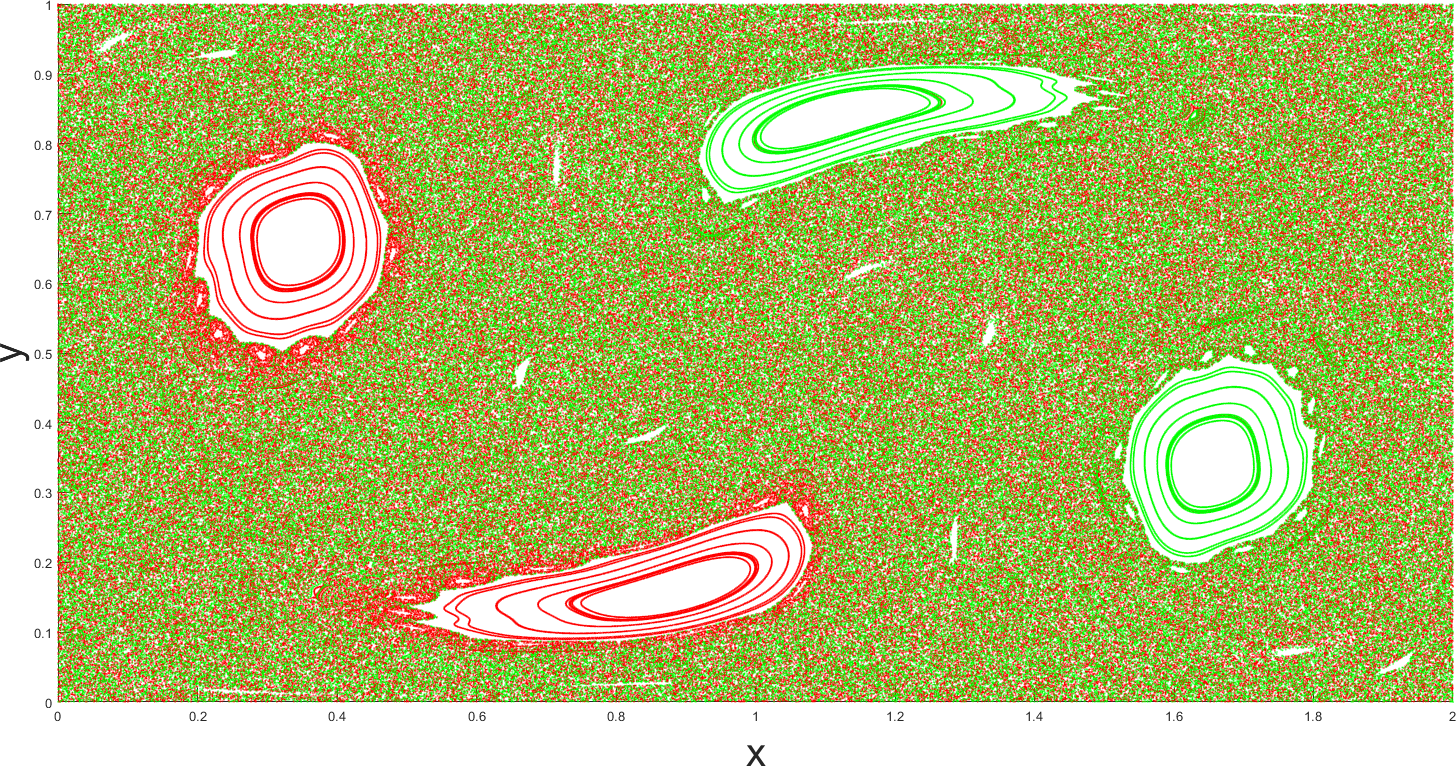}
\caption{}
\end{subfigure}
\begin{subfigure}[t]{0.33\textwidth}
\includegraphics[width=\textwidth]{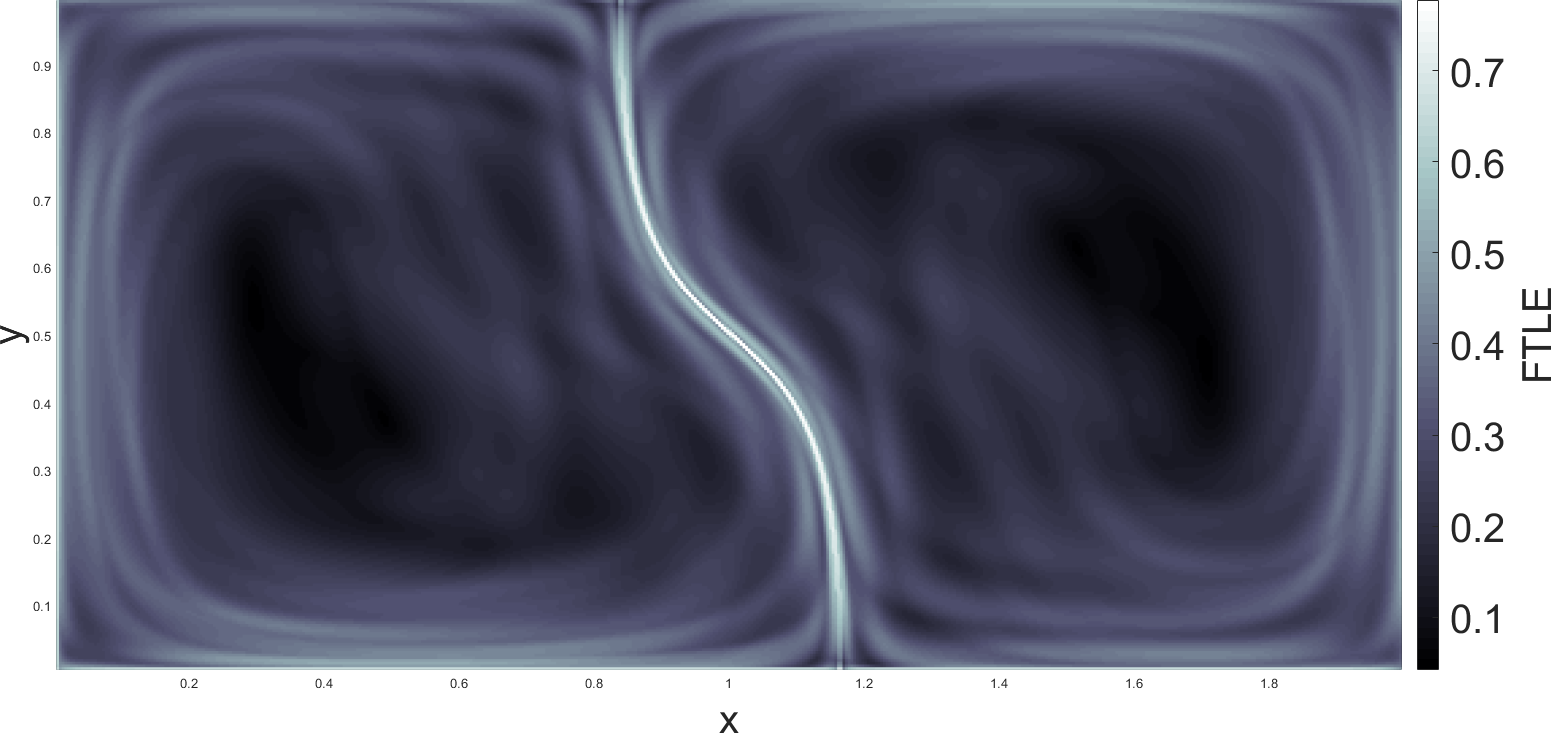}
\caption{}
\end{subfigure}\hfill
\begin{subfigure}[t]{0.33\textwidth}
\includegraphics[width=\textwidth]{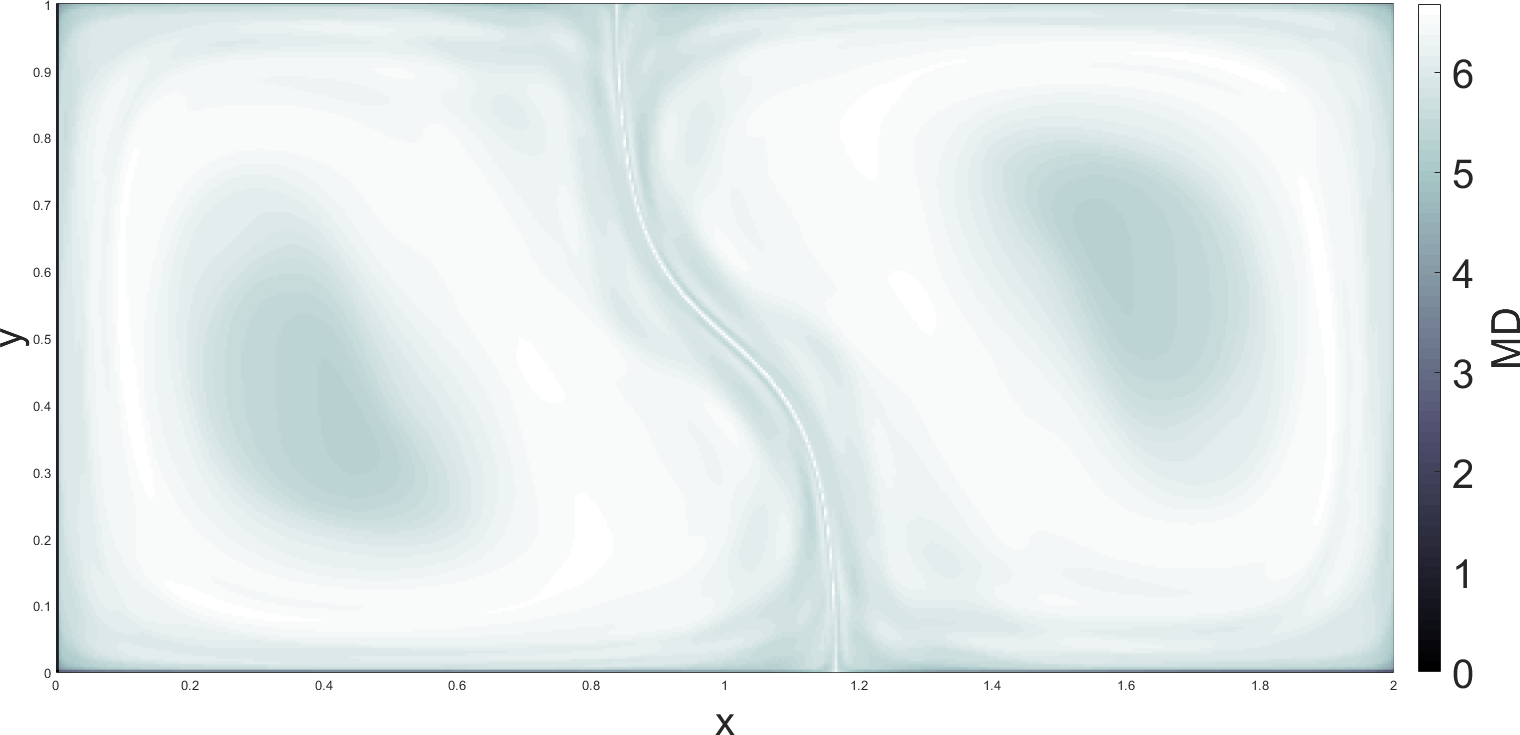}
\caption{}
\end{subfigure}\hfill
\begin{subfigure}[t]{0.31\textwidth}
\includegraphics[width=\textwidth]{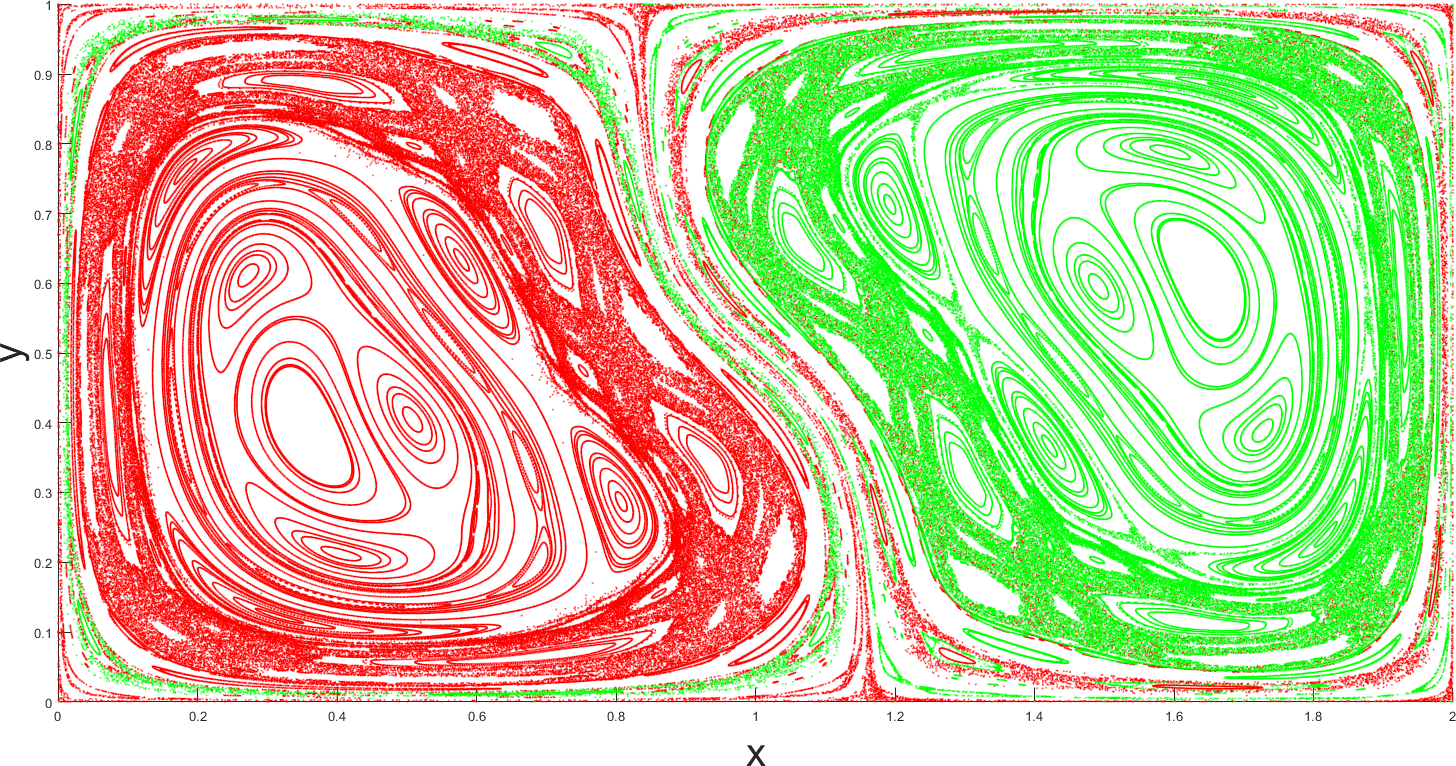}
\caption{}
\end{subfigure}
\caption{LDs, FTLEs, and KAM tori in the double-gyre such as in Equations \eqref{eqn:UPsi} and \eqref{eqn:timedep} with parameters $A = .1$ and $\epsilon = .5$. The first row [(a),(b),(c)] corresponds to $\omega = .5$, the second row [(d),(e),(f)] corresponds to $\omega = 1$, and the third row [(g),(h),(i)] corresponds to $\omega = 2$. The first column [(a),(d),(g)] contains summed forward and backward FTLE fields with $\tau = 15$, the second column [(b),(e),(h)] contains LD fields with $\tau = 15$, and the third column [(c),(f),(i)] contains Poincare Maps up to 4000 iterations of the period of perturbation.}
\end{figure}

\section{Discussion}

\hspace{\parindent} In this paper, the utility and computational considerations of Finite-Time Lyapunov Exponents and Lagrangian Descriptors were explored in detail, and many results were shown for a time-dependent double-gyre model. The double-gyre model is a toy model of a natural phenomenon common to large-scale ocean dynamics. \\

Some thought should be given to the choice of integration time, $\tau$, for both FTLEs and LDs. Throughout the paper, we have generally set $\tau = 15$. There is no general rule for choosing $\tau$, and the best choice is always problem-dependent. For periodically forced systems, a good place to start is often several multiples of the period of perturbation. For aperiodic systems, one might start with some advective time scale considering the size of the domain and magnitude of the velocity field. In general, as $\tau$ is increased, richer details of the underlying structures are revealed, meaning that the choice also depends on how much complexity the user wants to resolve. Ultimately, the process of choosing $\tau$ usually requires trial and error when attempting to reveal specific features. \\

When choosing to use FTLEs or LDs to reveal stable and unstable manifolds of hyperbolic trajectories, one consideration is our confidence in the method as a tool for revealing phase space structure. Branicki and Wiggins explored the accuracy of FTLE ridges for the purpose of revealing transport barriers (such as stable and unstable manifolds) in some depth \cite{branicki2010}. FTLE ridges are not material curves in general, though they may be ``close'' to one in the sense that flux across the ridge may be small or negligible with sufficiently large $\tau$, as discussed in \cite{shadden2005}. However, there are many known cases in which the stable and unstable manifolds of a system are known exactly, and yet FTLE fields are either flat, or their ridges do not correspond to stable and unstable manifolds. A few examples of such cases were shown in the Branicki and Wiggins work, including a simple time-dependent case of linear strain. For the example, the FTLE value of any particular initial condition, $\lambda_\tau(\vec{x},t_0)$, does not depend on $x$ or $y$. As shown in Figure \ref{fig:FTLEsingular}, the location of FTLE ridges can also be particularly sensitive to the choice in parameter, $\tau$. \\

In our results, KAM tori appear to be identified both in FTLEs and LDs. Some connection between invariant tori-like structures and FTLEs was shown in \cite{beron-vera2010}, where structures like KAM-tori are taken to be revealed by locally minimizing curves of FTLE fields. These ``trenches,'' however, appear to be subject to the same issues related to FTLE ridges that are discussed above. In general, one expects trajectories contained in invariant tori to experience minimal stretching compared to trajectories in chaotic regions of the domain, meaning trajectories within KAM tori will often have small FTLE values. The connection between LDs and invariant sets such as KAM tori, however, is more robust and has been rigorously considered by Lopesino et al. \cite{lopesino2017}. In principle, one can continue integration a bit longer, and compute the time average of $MD$ to extract an accurate approximation of KAM tori. This method has been used in Garc\'ia-Garrido et al. to extract invariant sets in 3D vector fields \cite{garcia-garrido2018}.\\

When identifying KAM tori, we observe that Poincare maps show more complex features of KAM tori than LDs or FTLEs in flows with periodic perturbations. However, while Poincare maps are useful for identifying KAM invariant tori in dynamical systems containing periodic perturbations, they are meaningless in aperiodic flows (there would be no ``natural'' frequency with which to infer recurrence in the first place). We may be interested, however, in finding similar ``islands'' in aperiodic systems, such as in many geophysical systems.  Lagrangian Descriptors and FTLEs may be more useful for identifying these features in aperiodic flows, since their computation does not depend on the existence of a characteristic frequency. Poincare Maps also require long integration lengths that are several orders of magnitudes longer than diagnostics such as FTLEs and LDs (usually several thousand multiples of the period of perturbation), in order to generate enough points to visibly differentiate between the broken and unbroken tori. \\

LDs are also exceptionally easy to calculate. An undergraduate student with minimal coding experience can be asked to compute the ``arc-length of trajectories,'' and will very likely return with satisfactory results. However, FTLEs require significantly more explanation, and are not always intuitive. One must first explain the meaning of Lyapunov exponents, reduce the concept to a finite-time case, calculate trajectories, differentiate, and then compute eigenvalues. This can be a difficult set of concepts for a new dynamical systems student to grasp. On the other hand, the conceptual simplicity of LDs is likely the reason they have begun to be accepted in new fields such as in chemical reaction dynamics \cite{katsanikas2020} \cite{naik2019}. \\

There are also fewer computational considerations for LD fields compared to FTLE fields. The sensitivity of FTLEs to grid spacing and integration time, for instance, requires the user to take more care when interpreting structure in FTLE fields. The choice of ``neighboring trajectories'' is also extremely important when computing FTLEs, and this fact can be easily neglected. For several years, the author made the mistake of computing FTLEs by generating a new set of initial conditions for each grid point, such as the problem illustrated in Figure \ref{fig:FTLEGrid}, which is not an uncommon mistake for students who are new to the dynamical systems approach to Lagrangian transport. \\

One might also like to know how FTLEs and LDs perform in higher dimensions. While FTLEs are commonly computed in three dimensions by constructing the appropriate three-dimensional deformation tensor, the identified structures suffer from similar limitations as they do in two dimensions \cite{branicki2010}. Meanwhile, the connection between Lagrangian Descriptors and phase-space structures in three and higher dimensions has been thoroughly detailed in a number of works \cite{garcia-garrido2018} \cite{lopesino2015} \cite{lopesino2017}  \cite{naik2019}. \\

While FTLEs have been used widely in literature to identify phase space structures in real systems such as oceanic flows, the concept of LDs is more recent, and the current extent of literature using LDs to analyze real flows is somewhat limited but growing rapidly. Future studies could see LDs being used more frequently to analyze these systems, considering their robust connection to phase space structures and their simplicity for new students to learn. 

\section*{Acknowledgments}
\hspace{\parindent} I would like to thank Steve Wiggins and Reza Malek-Madani for their tutoring in dynamical systems concepts, and for guidance related to this paper. Also, many thanks to the University of Bristol School of Mathematics for hosting during my visit in the summer of 2018, where the work began. I would also like to give special thanks to Kevin McIlhany for his many years of mentorship in CFD and other topics in physics and mathematics. 

\bibliographystyle{siam}
\bibliography{main.bib}
\end{document}